\documentclass[reprint,superscriptaddress,aps]{revtex4-2}
\usepackage{amsmath}
\usepackage{amssymb}
\usepackage{graphicx}
\usepackage{epstopdf}
\usepackage[colorlinks=true]{hyperref}
\usepackage{physics}
\usepackage{mathrsfs}
\usepackage{comment}

\renewcommand\({\begin{equation}}	
\renewcommand\){\end{equation}}
\renewcommand\[{\begin{eqnarray}}	
\renewcommand\]{\end{eqnarray}}
\newcommand{\al}[1]{\begin{aligned}#1\end{aligned}}
\usepackage[dvipsnames]{xcolor}

\begin{document}

\title{Spatially correlated classical and quantum noise in  driven qubits: \\ The good, the bad, and the ugly}

\author{Ji Zou}
\affiliation{Department of Physics, University of Basel, Klingelbergstrasse 82, 4056 Basel, Switzerland}
\author{Stefano Bosco}
\affiliation{Department of Physics, University of Basel, Klingelbergstrasse 82, 4056 Basel, Switzerland}
\author{Daniel Loss}
\affiliation{Department of Physics, University of Basel, Klingelbergstrasse 82, 4056 Basel, Switzerland}

\begin{abstract}
Correlated noise across multiple qubits poses a significant challenge for achieving scalable and fault-tolerant  quantum processors. Despite recent experimental efforts to quantify this noise in various qubit architectures, a comprehensive understanding of its role in qubit dynamics remains elusive. Here, we present an analytical study of the dynamics of driven qubits under spatially correlated noise, including both Markovian and non-Markovian noise. Surprisingly, we find that, while correlated classical noise only leads to correlated decoherence without increasing the quantum coherence in the system, the  correlated quantum noise can be exploited  to generate entanglement. 
{In particular, we reveal that, in the quantum limit, pure dephasing noise induces a coherent long-range two-qubit Ising interaction that correlates distant qubits.}
In contrast, for purely transverse noise  when qubits are subjected to coherent drives, the correlated quantum noise induces both coherent symmetric exchange and Dzyaloshinskii-Moriya interaction between the qubits, as well as correlated relaxation, both of which give rise to significant entanglement. Remarkably, in this case,  we uncover that the system exhibits distinct dynamical phases  in different parameter regimes. 
Finally, we reveal the impact of spatio-temporally correlated $1/f$ noise on the decoherence rate, and how its temporal correlations {restore} lost entanglement. 
{Our analysis not only offers critical insights into designing effective error mitigation strategies to reduce harmful effects of correlated noise, but also enables tailored protocols to leverage and harness noise-induced correlations for quantum information processing.
}
\end{abstract}

\date{\today}
\maketitle

\section{Introduction}
Quantum computers hold great promise for solving computational problems that are intractable for classical ones, due to their ability to exploit the quantum coherence of qubits~\cite{shor_1994,preskill2018quantum}. However, {quantum coherence is extremely fragile and noise } poses a major challenge {to quantum information processing}~\cite{schlosshauer2019quantum,nielsen}.
 A comprehensive understanding of the effects of noise is the first step towards the development of effective noise mitigation strategies, and  therefore, is crucial {to leverage} the full potential of {large-scale quantum processors}~\cite{suter2016colloquium,de2021materials}.
 
In the single-qubit scenario, the noise experienced by the qubit {is local and} can be characterized by a single noise spectrum~\cite{gardiner2004quantum,clerk2010introduction}. It has been experimentally measured in different quantum computation architectures, including spin qubits in semiconductors~\cite{chan2018assessment,kuhlmann2013charge}, nitrogen-vacancies~\cite{bar2012suppression,romach2015spectroscopy}, trapped ions~\cite{frey2017application,frey2020simultaneous}, and superconducting quantum circuits~\cite{bylander2011noise,yan2013rotating,quintana2017observation,yan2018distinguishing}, by employing  different  noise spectroscopy protocols. While the effect of single-qubit noise, in particular $1/f$ noise in solid state architectures, is still being actively investigated in order to develop better strategies for mitigating its impact~\cite{mutter2022fingerprints,mutter2023theory,gulacsi2023smoking,zhang2022predicting,paladino20141}, we have gained a relatively good understanding of its characteristics~\cite{divincenzo2005rigorous,khaetskii2002electron,bertet2005dephasing,coish2010free,coish2004hyperfine,martinis2003decoherence,schreier2008suppressing}.  Various techniques to address it in quantum computation have been proposed, including decoherence-free sweet spots~\cite{bosco2021hole,bosco2021fully,bosco2022hole,reed2016reduced,piot2022single,zhao2021practical}, quantum error correction codes~\cite{laflamme1996perfect,knill1997theory,ng2009fault,preskill2012sufficient}, dynamical decoupling~\cite{violaprl1999,khodjasteh2005fault,Paz_Silva_2016,cywinski2008enhance}, and optimal control methods~\cite{palao2002quantum,nielsen2006optimal,li2017hybrid,hansen2022implementation,hansen2021pulse,bosco2023phase,rimbach2022simple}, which  have shown promise in reducing the impact of single-qubit noise and improving the performance of quantum devices.

However, the presence of spatially correlated noise can limit the applicability of proposed protocols in multi-qubit settings.  For instance, major quantum error-correcting codes rely on {independently} detecting and correcting errors on individual qubits~\cite{laflamme1996perfect,knill1997theory,ng2009fault,preskill2012sufficient}. Correlated noise across multiple qubits impedes the effectiveness of these codes, leading to a higher probability of errors remaining undetected and reducing the performance of quantum systems~\cite{PhysRevA.57.120,hutter2014breakdown,klesse2005quantum,google2021exponential,wilen2021correlated}. { It is therefore crucial to better quantify and understand correlated noise}. This has stimulated  various theoretical proposals~\cite{paz2017multiqubit,szankowski2016spectroscopy,krzywda2019dynamical,rivas2015quantifying} and experimental works in the measurement of  correlated noise, for example, in architectures based on spin qubits~\cite{yoneda2022noise,rojas2023spatial} and superconducting qubits~\cite{von2020two}.
However, despite the  progress in quantifying  spatial noise correlations, a comprehensive understanding of how they affect the performance of multiqubit systems is still lacking.

On the other hand, while spatially correlated noise can have detrimental effects on quantum systems, it raises the question of whether the correlations stored in the noise, which are absent in the case of a single qubit, can be harnessed to {process} quantum information~\cite{aash_prl_2022,zou2022prb}. For instance, {correlated noise offers the intriguing possibility } to imprint the correlations onto a two-qubit system, effectively converting the correlation of noise into entanglement between the qubits. 
To develop effective strategies either for  mitigating correlated noise (suppressing the ``ugly" aspect) or for leveraging the correlations it stores (exploiting the ``good" aspect),  a comprehensive  understanding of the effects of correlated noise in multiqubit settings  is required.

In this work, we present a systematic theoretical investigation of the impact of spatially correlated noise on the dynamics of driven qubits with a focus on their entanglement, considering both temporally correlated and uncorrelated noise.
In particular, we find that operating the qubits at higher temperatures can help mitigate the  ``ugly" aspect of correlated noise and suppress the crosstalks between qubits. This unexpected reduction in crosstalk between qubits at warmer temperatures has been recently observed in an  experimental study on spin qubits~\cite{undseth2023hotter}.   
On the other hand, to exploit the ``good" aspect of correlated noise, such as the generation of substantial  long-lived entanglement, one needs to drive the qubits and operate the system at low temperatures (relative to the qubit energy).  We highlight that this entanglement generation process can be on-demand controlled by turning on and off the driving.

The present paper is organized as follows.  In Sec.~\ref{sec2}, we establish the foundation for our study. First, we introduce the model Hamiltonian in Sec.~\ref{sec21}. Next, we discuss the concept of local and spatially correlated noise spectral densities in Sec.~\ref{Sec_2.1},  distinguishing their classical and quantum components. Finally, we present a set of broadly applicable master equations for the driven dynamics of two-qubit systems in the presence of spatially correlated generic noise in Sec.~\ref{sec23}, which sets the stage for the following discussion.

In Sec.~\ref{sec_3}, we investigate spatially correlated  $1/f$ noise in {pure dephasing} dynamics without coherent drives. We pay particular attention to how the classical and quantum components participate in the two-qubit dynamics and whether they can be exploited to generate entanglement. We demonstrate that the classical correlations in the noise affect the dynamics through correlated pure dephasing, which modifies the dephasing rate but does not induce any coherence. In contrast, we reveal that correlated quantum noise  impacts the two-qubit dynamics through both {a noise-induced} coherent interaction between qubits and correlated pure-dephasing processes. Interestingly, the former allows for the conversion of noise correlation to entanglement, while the latter does not.

In Sec.~\ref{seciv}, we present an analytical study of the two-qubit dynamics under the influence of spatially-correlated Markovian transverse noise with coherent drives. Our analysis reveals that the quantum noise correlations induce a coherent symmetry exchange interaction, a Dzyaloshinskii–Moriya interaction between the two qubits, and a correlated decoherence process. We observe an intriguing interplay between these ingredients, resulting in distinct dynamical phases that can be achieved with different parameter values.   Surprisingly,   in contrast to  pure dephasing, we find that both coherent interactions and correlated decoherence   lead to a generation of  significant  entanglement, which can be potentially leveraged to implement two-qubit gates and other quantum information processing tasks.

Finally, in Sec.~\ref{secv},  we conduct an analytical investigation into the influence of correlated  $1/f$ noise on the dynamics of driven qubits. Our study shows that the non-Markovianity of the $1/f$ noise results in effective time-dependent  decoherence rate that exhibits temporary negative values for some time intervals~\cite{piilo2008non}. We focus on the classical and quantum correlated $1/f$ noise in Sec.~\ref{secv1} and~\ref{secv2}, respectively. We find that the classical spatially-correlated $1/f$ noise is still unable to generate any entanglement. However, the non-trivial temporal correlations of the noise can restore  the coherence lost in the environment. On the other hand, the non-Markovian nature of the quantum correlated $1/f$ noise  leads to a temporary decrease of the entanglement generated by the quantum noise.

\section{Theoretical Model} \label{sec2}
\subsection{Hamiltonian} \label{sec21}
\begin{figure}
	\centering\includegraphics[width=0.84\linewidth]{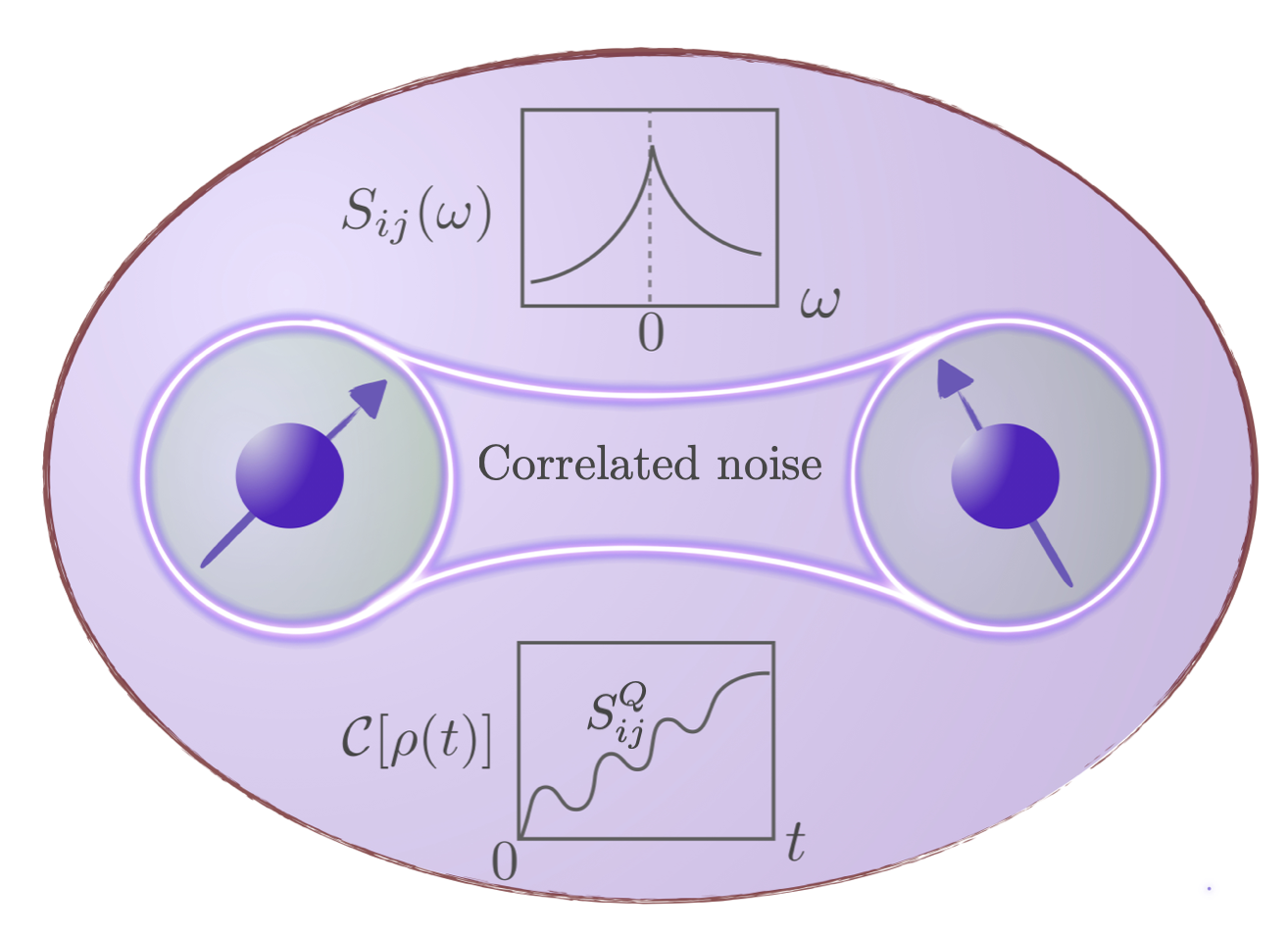}
	\caption{A schematic for two qubits in an environment experiencing both local and spatially correlated noise. The noise is quantified by the cross-noise power spectral densities $S_{ij}(\omega)$, with its positive and negative frequency parts measuring the ability of qubits to emit and absorb energy, respectively. The asymmetric nature of the noise spectral density, in the presence of quantum noise, is linked to the asymmetry between the absorption and emission processes. The quantum correlated noise can be harnessed {to generate  entanglement}.}
	\label{fig1}
	
\end{figure}

In this work, we analyse the dynamics of two qubits that are driven and are situated in the same environment, as depicted in Fig.~\ref{fig1}. The qubits are subjected to both local and spatially-correlated (non-local) noise, which can have either a classical or quantum nature and can be either temporally correlated (such as $1/f$) or uncorrelated (``Markovian").
 We describe the combined system with the following Hamiltonian: 
\(  H(t)=H_S+H_{\text{drive}}(t)+H_{\text{SE}}+H_E, \)
where $H_S$ is the Hamiltonian of the two qubits that are characterized by qubit-frequency splittings $\Delta_i$ ($i=1, 2$), and $H_{\text{drive}}(t)$ describes the time-dependent driving of the two qubits, $H_E$ is the Hamiltonian of the environment, and $H_{\text{SE}}$ describes the coupling between the two qubits and the environment. We assume the two qubits are driven coherently at frequency $\omega_{di}$ with the drive amplitude $\hbar\Omega_i$. We then consider the following Hamiltonian:
\( H_S+H_{\text{drive}}(t)= \sum_{i=1,2} \bigg[  \frac{\hbar \Delta_i}{2}\sigma_i^z +\hbar \Omega_i \cos(\omega_{di}t ) \sigma_i^x   \bigg],   \)
where $\sigma_i^{x,z}$ are the Pauli matrices for qubit $i$ and $\hbar$ is the Planck's constant. We describe the environment with \(H_E=\sum_{\vb k} \hbar \omega_{\vb k} b^\dagger_{\vb k} b_{\vb k},  \)
where $b_{\vb k}$ and $b^\dagger_{\vb k}$ are operators describing quasiparticles in the environment leading to the decoherence of the qubits, such as phonons in semiconductors~\cite{golovach2004phonon,bosco2022fully,kornich2014phonon} or magnons in hybrid systems~\cite{Daniel2013prx,Daniel2012prx,Fukami2021prx,hetenyi2022long,zou2022prb,zou2022domain,PhysRevB.101.014416,quantumvortex,Yqwinding}. While we leave the spectrum $\omega_{\vb k}$ unspecified which can be either  linear, for example, for acoustic phonons or quadratic for magnons, we specialize to single axis qubit-environment couplings: 
\(  H_{\text{SE}}= \sum_{i=1,2}\sigma_{i}^z E_i, \;\;\; \text{with}\;\;\; E_i \equiv g_{\vb k} e^{i\vb k\cdot \vb r_i}b_{\vb k}+\text{H.c.},   \)
implying that  pure-dephasing dynamics dominates  the decoherence process in the absence of coherent drives.
Here, $E_i$ are operators acting on the environment Hilbert space,  $\vb r_i$ is  the positions of the $i$th qubit, and $g_{\vb k}$ is the coupling strength.

When the system is subjected to coherent drives, we effectively rotate the quantization axis in a   frame rotating at the driving frequencies, where the qubits can exchange not only  information with the environment but also exchange  energy  by emitting and absorbing  quasiparticles. To illustrate this, we perform the unitary transformation  $R(t)=\exp(i\omega_{d1}\sigma_1^zt/2)\otimes \exp(i\omega_{d2}\sigma_2^zt/2)$.  The Hamiltonian in the rotating frame then is given by $\mathcal{H}=i\hbar \partial_t R R^\dagger +RHR^\dagger$. We note that the unitary operator $R(t)$ commutes with $H_{\text{SE}}$ and $H_E$, leaving them invariant in the rotating frame. After applying the rotating-wave approximation by neglecting counter-rotating terms that oscillate fast at frequency $2\omega_{di}$, the total Hamiltonian reads,
\( \mathcal{H}=\mathcal{H}_S +H_{\text{SE}}+H_{E},   \label{Eq:5}   \)
where the qubit Hamiltonian in the new rotating frame is $\mathcal{H}_S=\sum_{i=1,2} \hbar\big( \delta_i\sigma_i^z +\Omega_i\sigma_i^x  \big)/2$ with the detuning $\delta_{i}=\Delta_i-\omega_{di}$. When the qubits are driven at resonance  $\omega_{di}=\Delta_i$,  we arrive at:
\( \mathcal{H}_S= \frac{\hbar\Omega}{2}  \sum_{i}\hat{\sigma}_i^z,\;\;\; \text{and}\;\;\; H_{\text{SE}}=-\sum_i \hat{\sigma}_i^x E_i.  \)
Here, we have rotated the axis in the spin space, $\mathcal{H}_S\rightarrow R_y \mathcal{H}_S R_y^\dagger$ with $R_y=\text{exp}[i(\pi/2)\sigma_y/2]$, such that the qubit quantization axis is aligned with the $z$ axis, and we label this new basis with Pauli matrices $\hat{\sigma}_i$. We  also assume here that the driving strength is equal for both qubits, denoting it as $\Omega\equiv\Omega_1=\Omega_2$. We remark that, in this scenario, the relaxation dynamics dominates the decoherence process, which has been exploited for noise-spectroscopy applications to extract noise spectra near frequency $\Omega$~\cite{von2020two}. We focus our analysis in this work on the effects of correlated noise in two fundamental scenarios: one in the absence of coherent drives where pure-dephasing noise dominates, and the other in the presence of resonant drive where the transverse noise is prominent.

\subsection{Distinguishing classical from quantum noise in local and spatially correlated noise} \label{Sec_2.1}
We  described the qubit Hamiltonian under external coherent drives in the preceding section.
Here, we discuss the noise experienced by the qubits due to their coupling to the environment, focusing on both local and spatially correlated noise while distinguishing their classical and quantum natures.  We define the usual two-point noise correlation function in time domain~\cite{clerk2010introduction}:
\( S_{ij}(t)\equiv \langle E_i(t)E_j(0)\rangle,  \)
where $E(t)\equiv e^{iH_Et/\hbar} Ee^{-iH_Et/\hbar} $ and  $\langle\mathcal{O} \rangle\equiv \tr (\rho_B \mathcal{O})$ with the thermal state $\rho_B=e^{-\beta H_E}/\tr[\exp(-\beta H_E)]$ and $\beta=1/k_BT$, where $T$ is the temperature and $k_B$ is the Boltzmann constant. The noise power spectral density is given by the Fourier transformation of the correlation function:
\( S_{ij}(\omega)=\int^\infty_{-\infty} dt\; e^{i\omega t}  S_{ij}(t).  \)
Here, $S_{ii} (\omega)$ with $i=\{1,2\}$ is the auto power spectral density standing for the local noise, whereas $S_{ij}(\omega)$ with $i\neq j$ is the cross power spectral density representing spatially correlated noise. We note that $S_{ij}(\omega)=S_{ji}^*(\omega)$, indicating that the local noise spectral density is a real-valued function while the correlated noise can be complex-valued. This is also clear from the explicit expression of the noise spectral density: 
\( \al{S_{ij}(\omega)= & 2\pi \sum_{\vb k} |g_{\vb k}|^2 e^{-i\vb k\cdot (\vb r_j- \vb r_i)} \big[ n_B(\omega_{\vb k}) +1   \big] \delta(\omega-\omega_{\vb k})  \\
    &+ 2\pi \sum_{\vb k} |g_{\vb k}|^2 e^{i\vb k\cdot (\vb r_j- \vb r_i)} n_B(\omega_{\vb k})  \delta(\omega+\omega_{\vb k}),  }  \label{Eq:9}  \)
where $n_B(\omega)=1/(e^{\beta \hbar \omega}-1)$ is the Bose-Einstein distribution and the spatial vector $\vb r_j-\vb r_i$ connects the positions of qubit $j$ and $i$. The  cross power spectral density $S_{12}(\omega)$ is real when the spectrum of the environment is symmetric in momentum $\omega_{\vb k}=\omega_{-\vb k}$, whereas it is complex for general asymmetric $\omega_{\vb k}$, for example in inversion-asymmetric environments. 

It is worth noting that the noise power spectral density~\eqref{Eq:9}  is generally asymmetric in frequency, $|S_{ij}(\omega)|\neq |S_{ji}(-\omega)|$. Its positive- and negative-frequency components are linked through the Boltzmann factor, $S_{ij}(\omega)=e^{\beta\hbar \omega}S_{ji}(-\omega)$, which reflects the quantum nature of the noise, as indicated by the non-zero commutator $[E_1(t), E_2]\neq 0$.
One can interpret the positive-frequency part $S_{ij}(\omega>0)$ as a measure of the ability of  qubits to emit energy, and the negative-frequency part $S_{ij}(\omega<0)$  as a measure of the ability of  qubits to absorb energy. To differentiate between quantum and classical noise, we introduce the symmetrized and antisymmetrized noise spectral densities~\cite{engel2004asymmetric,clerk2010introduction}:
 \( S^C_{ij}(\omega)\!\! =\! \!\frac{S_{ij}(\omega)\! +\! S_{ji}(-\omega)  }{2},\; S^Q_{ij}(\omega)\!\!=\! \!\frac{S_{ij}(\omega) \!-\! S_{ji}(-\omega)  }{2},  \label{eq:cqnoise} \)
which are linked to each other through $S^C_{ij}(\omega)=\coth(\beta \hbar\omega/2) S^Q_{ij}(\omega)$, as dictated by the fluctuation-dissipation theorem~\cite{landauvol5}. This {distinction enables} a better understanding of how different types of noise affect the behavior of multiqubit systems.

\begin{figure}
	\centering\includegraphics[width=\linewidth]{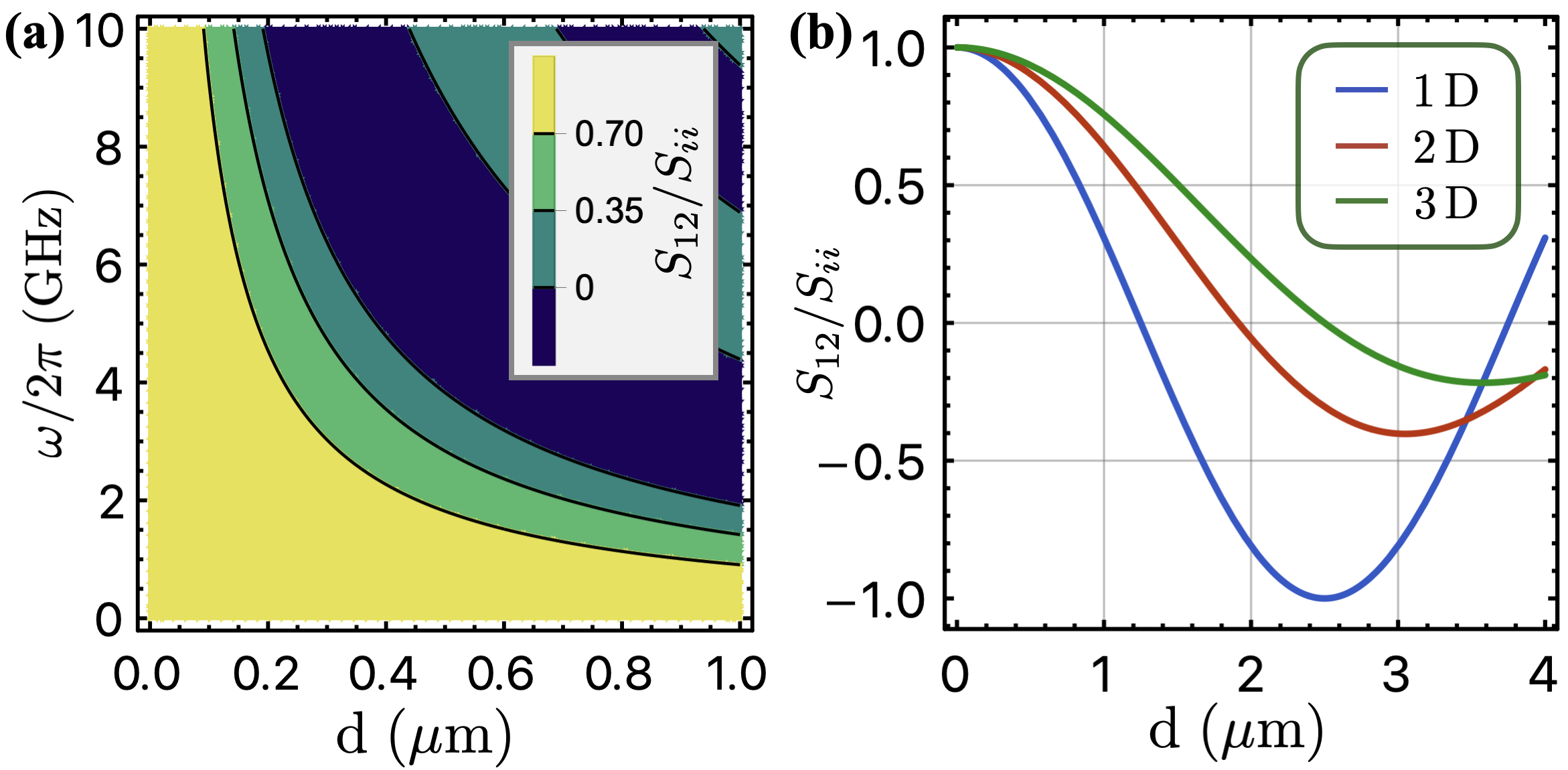}
	\caption{Correlated noise. (a) Density plot of the ratio between the spatially correlated noise $S_{12}$ and the local noise $S_{ii}$ as a function of frequency $\omega$ and the distance $d$ between two qubits in two dimensions.  The plot shows that for qubit frequencies in the few GHz regime, the correlated noise is as strong as the local one when two qubits sit within micrometers, and it oscillates and decays to zero as the distance increases. (b) The ratio $S_{12}/S_{ii}$ as a function of qubit distance in different dimensions, where we have taken $\omega/2\pi=1\ \text{GHz}$. The correlated noise exhibits similar behavior in different dimensions, and obeys the constraint $S_{12}\leq S_{ii}$.  In both plots, we used $\omega_\textbf{k}=c_s |\textbf{k}|$ and $c_s=5\, \text{km/s}$. }
	\label{fig2}
\end{figure}

When operating qubits at high temperatures, such as in the case of spin qubits that can be operated at a few Kelvin~\cite{yang2020operation,camenzind2022hole,petit2020universal,petit2022design}, the classical limit $k_BT\gg \hbar \omega$ applies, and classical fluctuations dominate over quantum noise.  In this regime, the operators $E_i(t)$ can be treated classically $[E_i(t), E_j(0)]=0$, resulting in a symmetric noise spectral density with a vanishing antisymmetric part, $S^Q=0$. In contrast, in the quantum regime where $k_BT\lesssim \hbar \omega$, the quantum noise is comparable to classical noise, with $S^Q\approx S^C$, and therefore cannot be neglected.

We point out that there is  a constraint for the spatially correlated noise, $|S_{12}^C(\omega)|^2 \leq  S^C_{11}(\omega) S^C_{22}(\omega)$~\cite{clerk2010introduction}, implying that the  nonlocal noise is inherently bounded by the local one. This condition is closely related to the thermodynamic stability of the environment~\cite{zou2022prb}. 
To illustrate this condition explicitly, we consider a concrete example, where the environment has a linear spectrum $\omega_{\vb k}=c_s |\vb k|$. This case describes for example  acoustic phonons with sound velocity $c_s$. The spatially correlated noise is related to the local noise in a 2D architecture through the following equation:
\(S_{12}(\omega)=J_0(\omega d/c_s) S_{ii}(\omega),\) where $|J_0(x)|\leq 1$ is the Bessel functions of the first kind and $d=|\vb r_1-\vb r_2|$ is the distance between two qubits, as shown in Fig.~\ref{fig2} (a). At large distances, the correlated noise decays as $J_0(\omega d/c_s) \sim 1/\sqrt{d}$.   We will assume the two qubits to be identical and thus they experience the same local noise $S_{11}=S_{22}$ throughout our discussion. The constraint $|S_{12}|\leq S_{ii}$ also holds in other dimensions, as shown in Fig.~\ref{fig2} (b) where the ratio $S_{12}/S_{ii}$ is shown as a function of the separation between the two qubits in different dimensions. Exact relations  are provided in Appendix~\ref{appendix1}. 

 We remark that, when the two qubits are sitting within a few hundred nanometers, the spatially correlated noise is {comparable to} the local noise for qubit frequencies in the gigahertz range, as illustrated in Fig.~\ref{fig2} (a), and assuming the sound velocity to be $c_s\sim 5\,\text{km}/\text{s}$ in silicon. The correlated noise exhibits oscillatory behavior and gradually decays to zero as the qubit separation $d$ increases.
  As shown in Fig.~\ref{fig2} (b) where we take the frequency to be $\omega/2\pi=1\, \text{GHz}$, the spatially correlated noise is always comparable with the local noise for $d<1\,\mu\text{m}$ in different dimensions.

\subsection{Time convolutionless master equation}\label{sec23}
We now investigate the role of spatially correlated classical and quantum noise in the dynamics of the two-qubit system by deriving a master equation for the reduced density matrix  $\rho(t)$, obtained by tracing out the environment from the total density matrix $\rho_{\text{tot}}(t)$. To this end, we adopt a standard time-convolutionless (TCL) master equation approach~\cite{Heinz} and assume that the qubit-environment interaction is weak enough to truncate the TCL generator at the second order.  Leaving the detailed derivation to the Appendix~\ref{appendix2},  here we present the TCL master equations for the two-qubit system subject to correlated noise, without and with a resonant drive, respectively. In particular, we  separate the quantum and classical noise, enabling us to clearly identify their respective contributions to the qubit dynamics. Our results allow us to explicitly calculate and explore the effects of correlated noise.

\textbf{Pure-dephasing noise.}
In the absence of coherent driving, the qubit dynamics is purely determined by dephasing. As detailed in Appendix~\ref{appendixb2}, the TCL master equation for the two-qubit system, in the interaction picture, takes the form of
\( \dot{\rho}(t)=-i [ H_z(t), \rho(t)  ] +\sum_{i,j\in\{1,2\}} \mathcal{L}^z_{ij}(t)\rho(t),  \)
where  the correlated-noise-induced   coherent interaction between qubits is Ising-like: \(H_z(t)=\mathcal{J}^z(t)\sigma_1^z\sigma_2^z.\)
 The superoperators $\mathcal{L}^z_{ij}(t)$ are defined by
\( \mathcal{L}^z_{ij}(t)\rho= \gamma^z_{ij}(t)\Big[ \sigma_j^z \rho \sigma_i^z -\frac{1}{2}\{  \sigma_i^z\sigma_j^z,\rho    \}   \Big],   \)
with the anticommutator $\{A,B\}\equiv AB+BA$. Here, $\gamma^z_{ii}(t)$ stands for the standard  local dephasing which is time-dependent in general, whereas $\gamma^z_{12}(t)$ represents correlated  dephasing originated from the spatially correlated noise. 
The time-dependent coherent coupling parameter is given by $\mathcal{J}^z(t)=\int^t_0ds\big[ \mathcal{G}^R_{12}(s) +\mathcal{G}^R_{21}(s)  \big]/2\hbar^2$, quantifying the retarded interaction mediated by the environment, where $\mathcal{G}_{12}^R(t)=-i\Theta(t)\langle [E_1(t), E_2]  \rangle$ is the standard retarded Green's function. To clarify its relation to the correlated noise, it is insightful to recast it into the following form: 
\( \mathcal{J}^z(t) = \frac{4}{\hbar^2}\int^\infty_0 \frac{d\omega}{2\pi}\Re[ S^Q_{12}(\omega)] F_c(\omega,t), \label{eq14}   \)
with a filter function $F_c(\omega,t)=[\cos(\omega t)-1]/\omega$. This coherent Ising interaction is solely determined by the correlated quantum  noise.  One can also infer this conclusion from the fact that it is dictated by the retarded Green's function which vanishes when the noise operators $E_i$ commute, e.g.  if they are  classical variables.

Similarly, to investigate the effects of correlated classical and quantum noise on the dephasing, we  write the dephasing rate as 
\(  \gamma^z_{ij} (t) = \frac{4}{\hbar^2} \int^\infty_0 \frac{d\omega}{2\pi} F_s(\omega, t) \big\{ \Re[S^C_{ij}(\omega)]   +i \Im[S^Q_{ij}(\omega)] \big\}, \label{eq15}   \)
with the filter function  $F_s(\omega, t)=\sin\omega t/\omega$, which is peaked at zero frequency and approaches a delta function at large time $F_s(\omega, t\rightarrow \infty)=\pi \delta(\omega)$.  We observe that the local dephasing rate $\gamma^z_{ii}$ is determined solely by the classical noise $S^C_{ii}$ (where we recall that the local noise spectral densities $S_{ii}^{C,Q}$ are always real), whereas the correlated dephasing $\gamma_{12}^z$ has contributions from both classical and quantum spatially correlated noise. In contrast, we stress again that the coherent coupling is not affected by classical noise. Furthermore, it is worth noting that correlated quantum noise only influences the dephasing process when the environment has an asymmetric spectrum such that $\Im[S^Q_{12}]\neq 0$. Otherwise, $\gamma^z_{12}(t)$ is exclusively determined by the correlated classical  noise $S^C_{12}(\omega)$.

\textbf{Pure-transverse noise.}
In the presence of  resonant transverse drives, the combined system is governed by the Hamiltonian~\eqref{Eq:5} where the qubits experience pure transverse noise. The TCL master equation of the two-qubit system is shown to be the following form 
\( \dot{\rho}(t)=-i [ H_{xy}(t)  ,\rho(t)] +\sum_{i,j\in\{1,2\}} \mathcal{L}_{ij}(t)\rho(t), \label{eq:mastereq}   \)
where the coherent interaction is \(H_{xy}(t) =~\mathcal{J}(t) \hat{\sigma}_1^+\hat{\sigma}_2^-+~\text{H.c.}, \label{eq:hamiltonian}   \) 
and the superoperators are given by 
\( \al{ \mathcal{L}_{ij}(t)\rho &= \gamma_{ij}^\downarrow(t)\Big[ \hat{\sigma}_j^- \rho \hat{\sigma}_i^+ -\frac{1}{2}\{ \hat{ \sigma}_i^+\hat{\sigma}_j^-,\rho    \}   \Big]   \\
   &\; +  \gamma_{ij}^\uparrow(t)\Big[ \hat{\sigma}_j^+ \rho \hat{\sigma}_i^- -\frac{1}{2}\{  \hat{\sigma}_i^-\hat{\sigma}_j^+,\rho    \}   \Big] . } \label{eq:lindblad}  \)
   Here $\gamma_{ij}^{\downarrow}$ 
   and $\gamma_{ij}^\uparrow$ 
   describe time-dependent (local and correlated) decay and absorption rates, respectively. Detailed derivations are provided in Appendix~\ref{appendixb3}.
   Similar to the pure dephasing case, the coupling parameter $\mathcal{J}(t)$ is also fully determined by the quantum noise,  and can be  expressed in terms of the retarded Green's functions or correlated noise spectral densities as  \( \al{  \mathcal{J}(t)\!\! &=\!\! \frac{1}{2\hbar^2}  \int^t_0ds \big[ \mathcal{G}^R_{12} e^{i\Omega s} +\mathcal{G}^R_{21}(s)e^{-i\Omega s}  \big]  \\
    &=\!\! \frac{2}{\hbar^2} \!\!\int^\infty_0\!\!\! \frac{d\omega}{2\pi} \!\Big[\! S^Q_{12}(\omega) F_c(\omega-\Omega, t)\! +\! S^Q_{21}(\omega) F_c(\omega+\Omega, t)\! \Big] ,  }  \label{eq18} \)
    with qubit energy splitting $\Omega$.
We stress that, in contrast to the Ising coupling $\mathcal{J}^z(t)$ which is always real, $\mathcal{J}(t)$ is complex in general. 
   
   The local and correlated decay processes are induced by both classical and quantum noise, $ \gamma_{ij}^\downarrow(t) = \gamma^{C}_{ij}(t,\Omega)+ \gamma^{Q}_{ij}(t,\Omega)$, with classical and quantum contributions being
   \(   \al{\gamma^{C}_{ij}(t,\Omega)\!\!=\!\!\frac{2}{\hbar^2}\!\! \int^\infty_0\!\! \frac{d\omega}{2\pi}\! \Big[ \! S^C_{ij}(\omega) F_s(\omega \!\!-\!\!\Omega,t) \!\! +\!\!S^C_{ji}(\omega) F_s(\omega\!\!+\!\!\Omega,t) \!  \Big],  \\
               \gamma^{Q}_{ij}(t,\Omega)\!\!=\!\!\frac{2}{\hbar^2}\!\! \int^\infty_0\!\! \frac{d\omega}{2\pi} \!\Big[\! S^Q_{ij}(\omega) F_s(\omega\!\!-\!\!\Omega,t)\!\! -\!\! S^Q_{ji}(\omega) F_s(\omega\!\!+\!\!\Omega,t) \!  \Big].  }   \label{eq19} \)
  Similarly, the local and correlated absorption rates are  given by $\gamma_{ij}^\uparrow(t)= \gamma^{C}_{ij}(t,-\Omega)+ \gamma^{Q}_{ij}(t,-\Omega)$. Notably, unlike the local pure dephasing rate that is solely determined by classical noise, $\gamma_{ii}^{\uparrow,\downarrow}$ depends on both classical and quantum noise. Moreover, the correlated quantum noise is always present in the correlated decay and absorption rates, regardless of the symmetry of the spectrum $\omega_{\vb k}$.
   We emphasize that these processes exhibit high sensitivity to the noise spectra within a frequency window of approximately $1/t$ centered on the qubit splitting $\pm \Omega$. Therefore, for sufficiently long time evolution where $t\gg 1/\Omega$, we can consider exclusively the contribution of the spectra at $\omega=\pm \Omega$, and  approximate the rates as
  \( \gamma^\downarrow_{ij}\!=\!\frac{1}{\hbar^2} \big[  S^C_{ij}(\Omega) \!+\!  S^Q_{ij}(\Omega)  \big],\;\;\; \gamma^\uparrow_{ij}\!=\!\frac{1}{\hbar^2} \big[  S^C_{ji}(\Omega) \!-\!  S^Q_{ji}(\Omega)  \big] , \label{eq:markovianrate} \)
from where we observe that the asymmetry between the absorption and emission is caused by quantum noise. We  also obtain the standard detailed balance condition $\gamma^\downarrow_{ij}=e^{\beta \hbar \Omega}\gamma^\uparrow_{ji}$, by invoking the fluctuation-dissipation theorem.

The TCL master equations presented here for the two-qubit system show clearly the dependence of qubit dynamics on the quantum and classical components of local and spatially correlated noise. Our formalism is general, and it does not require any microscopic understanding of the spatial and temporal correlations within the environment, as long as the noise is weak enough to justify the truncation of the TCL generator at the second order. It can be employed to describe challenging cases such as long-ranged classical or quantum non-Markovian noise. Furthermore, this approach can be extended straightforwardly to multiple qubits. By separating the contributions of classical and quantum noise to the qubit dynamics, this scheme provides a foundation for our investigation of the impact of generic noise on multiqubit dynamics.

\section{ Pure dephasing process with correlated $1/f$ noise} \label{sec_3}

In this section, we utilize the formalism presented above to investigate the impact of spatially correlated classical and quantum noise on the dephasing of two qubits. Specifically, we focus on a noise spectral function with a $1/f$ frequency dependence, which is common in various quantum computing architectures~\cite{paladino20141}, including superconducting qubits and semiconducting devices. We stress that our approach is straightforwardly extended to  other noise spectra.
 Let us consider a  local classical  noise spectral density:
\(  S^C_{ii}(\omega)=\begin{cases} 2\pi \sigma^2/|\omega|, & \;\;\; \text{if}\; |\omega|>\omega_l, \\
                                                \;  \;\;0, & \;\;\; \text{otherwise},\end{cases}     \)
where $\sigma$ is the standard derivation of the noise and  $\omega_l$ stands for the low frequency cutoff that is set by the measurement time. The timescales that we investigate are  much shorter than this time, $t\ll \omega_l^{-1}$. As we aim to study the effect of the correlated noise, we consider  two qubits positioned  within the range of micrometers,
with correlated noise being comparable with the local one, $S_{12}^{C,Q}(\omega)\approx e^{i\theta}S^{C,Q}_{ii}(\omega)$, where $\theta$ is the phase of the correlated noise spectral density that characterizes its complex nature~\cite{correlated_noise_phase}. We point  that in our study, we assume that the distance between the two qubits is greater than the typical confinement lengths (10-50 nm for spin qubits). This ensures that the direct exchange interaction is suppressed, allowing us to focus specifically on the effects induced by correlated noise.

To examine how  the classical and quantum components of the spatially correlated noise determines the two-qubit dynamics, we consider the quantum regime, where both quantum and classical noises are present and $S^Q\approx S^C$. In this scenario,  the coherent coupling and the local dephasing rate $\gamma^z\equiv \gamma^z_{ii}$  are given by
\( \mathcal{J}^z(t)=- \frac{2\pi \sigma^2 \cos\theta}{\hbar^2}t,\;\; \gamma^z(t)=\frac{4\sigma^2t}{\hbar^2}\big[ 1- \text{Ci}(\omega_l t) \big], \label{eq22} \)
as detailed in Appendix~\ref{appendixc},
and the correlated dephasing rate is given by $\gamma^z_{12}=e^{i\theta}\gamma^z$ whose real and imaginary parts are rooted in classical and quantum correlated noise, respectively.  Here, $\text{Ci}(x)\equiv -\int^\infty_xd\tau\, \cos(\tau)/\tau$ is the cosine integral function.

\begin{figure}
	\centering\includegraphics[width=\linewidth]{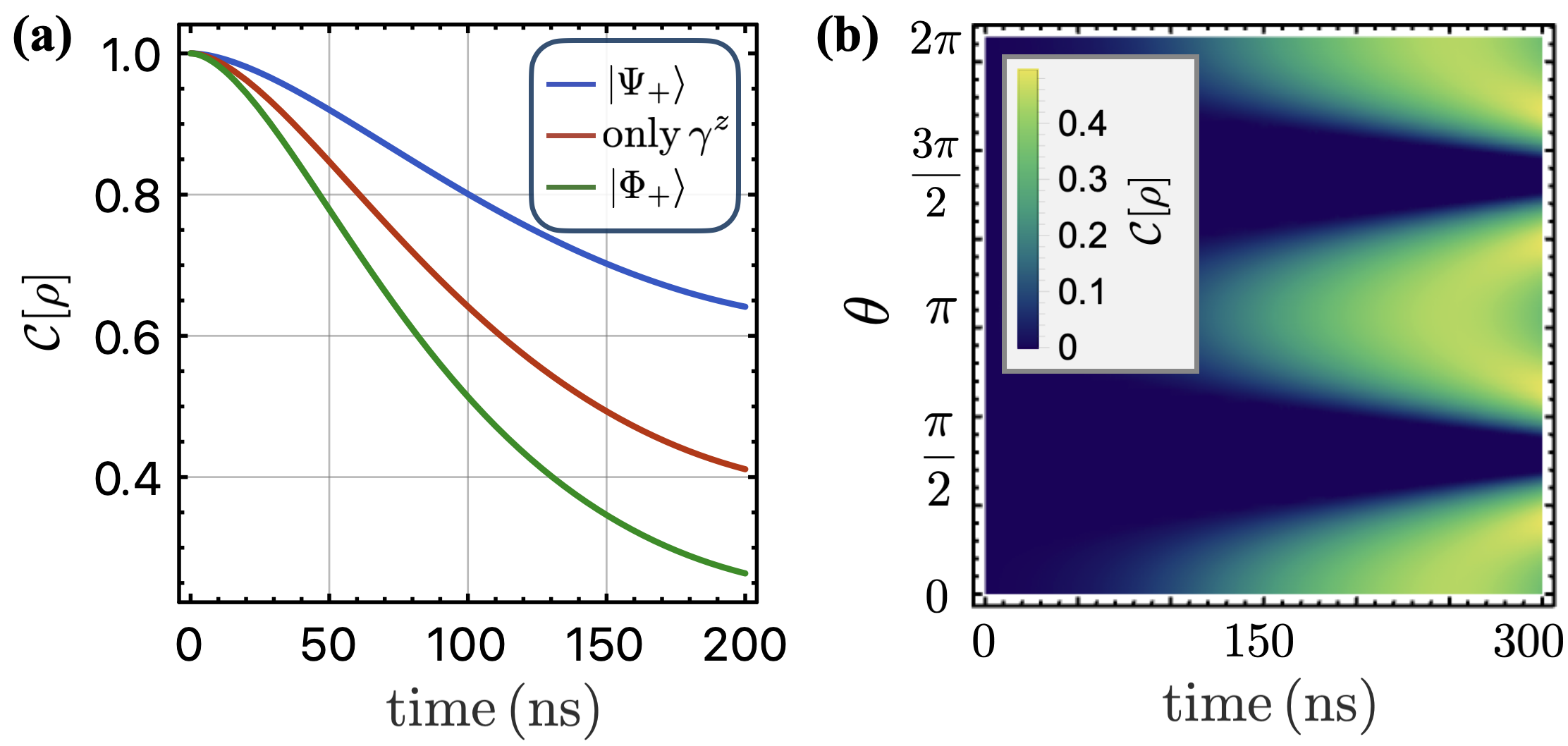}
	\caption{ Dephasing of two qubits subjected to spatially correlated noise. (a) Entanglement decay as a function of time for different initial states. The red curve corresponds to the case where only local noise is present, showing that both scenarios [with initial states being Bell states $\ket{\Psi_+}= ( \ket{\uparrow\downarrow} +\ket{\downarrow\uparrow} )/\sqrt{2}$ and  $\ket{\Phi_+}=( \ket{\uparrow\uparrow} +\ket{\downarrow\downarrow} )/\sqrt{2}$\,] decay at the same rates. The blue and green curves show the entanglement decay with initial states $\ket{\Psi_+}$ and $\ket{\Phi_+}$, respectively, in the presence of correlated classical noise. We set the quantum noise to zero and choose a phase of $\theta=\pi/3$ for the correlated noise power spectral density $S_{12}(\omega)$.  (b) Density plot of the entanglement as a function of time and the angle $\theta$ with initial state $\ket{++}$ ($\ket{+}$ is defined by $\sigma^x\ket{+}=\ket{+}$). The correlated quantum noise is real and enters the qubits dynamics through the coherent coupling when $\theta=0, \pi$, and is purely imaginary and enters the dynamics through the correlated dephasing when $\theta=\pi/2, 3\pi/2$. The correlated quantum noise does not generate entanglement through the correlated dephasing, but it generates it via the noise-induced Ising coherent coupling $\mathcal{J}^z$. Parameters used in the plot: $\hbar/\sigma=500\,\text{ns}$ and $\omega_l/2\pi=1\,\text{MHz}$. }
	\label{fig3}
\end{figure}

It is convenient to work in the basis $\{\ket{a}\}= \{ \ket{\uparrow\uparrow},\ket{\uparrow\downarrow},\ket{\downarrow\uparrow},\ket{\downarrow\downarrow} \}$, where the density matrix elements, denoted as $\rho=G_{ab}\ket{a}\bra{b}$, are all decoupled from each other. While the diagonal elements $G_{aa}$ remain  constant in the pure-dephasing dynamics, the off-diagonal components depend non-trivially on the correlated noise. By solving the TCL master equation analytically (see Appendix~\ref{appendixc} for details), we find that the classical component of the correlated  noise can only reduce or enhance the dephasing rate  caused by the local classical noise, without increasing the coherence in the two-qubit system. For instance, $G_{23}$ and $G_{14}$ are given by $G_{23}(t)  = G_{23}(0)\exp[-4(1-\cos\theta) \Gamma^z(t)  ]$ and  $G_{14}(t)  = G_{14}(0)\exp[-4(1+\cos\theta) \Gamma^z(t)  ]$ with 
\(\Gamma^z(t)\equiv \int^t_0ds\, \gamma^z(s) \approx \frac{\sigma^2 t^2}{\hbar^2} \big[  3-2\gamma -2 \ln (\omega_l t)\big],  \)
suggesting a Gaussian decay of the coherence with a logarithmic correction.  Here, $\gamma$ is Euler's constant.  

We are now ready to investigate how the correlated noise affects the dynamics of two-qubit entanglement. There are different measures of entanglement. For example, the singlet fidelity of the corresponding Werner state~\cite{werner1989quantum} of an arbitrary mixed state provides a lower bound for the entanglement of formation~\cite{burkard2003lower,bennett1996mixed}, as detailed in Appendix~\ref{appendixd1}. In this work, we adopt the two-qubit concurrence as a measure of entanglement~\cite{PhysRevLett.80.2245}, which is also summarized in Appendix~\ref{appendixd1}.
In Fig.~\ref{fig3} (a), we illustrate the entanglement decay of the two-qubit system with the initial states being Bell states $\ket{\Psi_+}\equiv ( \ket{\uparrow\downarrow} +\ket{\downarrow\uparrow} )/\sqrt{2}$ and $\ket{\Phi_+}\equiv ( \ket{\uparrow\uparrow} +\ket{\downarrow\downarrow} )/\sqrt{2}$, respectively. The red curve represents the entanglement decay when only local noise is present, where both states decay with the same rates. When the correlated classical noise is present (we set the quantum noise to zero and  $\theta=\pi/3$ ), both states still decay but with different rates, corresponding to the blue and green curve in Fig.~\ref{fig3} (a).

The quantum component of the correlated noise affects the two-qubit dynamics through both the coherent Ising interaction, which is governed by the real part of $S_{12}^Q$, and the correlated dephasing, which is linked to the imaginary part of $S_{12}^Q$. Interestingly, we find that only the real part of the quantum correlated noise leads to an increase in the entanglement between the two qubits. To illustrate this effect, we consider an initial state $\ket{++}$, where $\ket{+}$ is defined by $\sigma^x\ket{+}=\ket{+}$. We examine the entanglement dynamics as a function of time and phase $\theta$. When the spectral density is real ($\theta=0, \pi$), we observe a sizable increase in entanglement, while it remains at zero when the spectral density is purely imaginary ($\theta=\pi/2, 3\pi/2$), as shown in Figure~\ref{fig3} (b).

Our findings indicate that to harness the correlations encoded in the pure-dephasing noise, qubits must be operated at low temperatures to ensure the presence of quantum correlated noise and in an inversion symmetric environment to obtain a real quantum noise spectral density. Conversely, if one wishes to prevent undesired entanglement between the qubits, operating the qubits at higher temperatures to favor classical correlated noise or breaking the symmetry of the environment at low temperatures to suppress entanglement generation can be effective strategies.
We notice that a very recent experiment~\cite{undseth2023hotter} has observed an unexpected significant reduction in crosstalk between spin qubits at higher temperatures in semiconductors.

\section{ Driven spin qubits subject to correlated Markovian noise} \label{seciv}
In this section, we investigate the dynamics of two qubits when they are resonantly driven. This case is governed by Hamiltonian~\eqref{Eq:5}, and the reduced dynamics is described by the master equation~\eqref{eq:mastereq}. In particular, we focus on the temporally uncorrelated (Markovian) noise, because this model accurately describes  noise with generic power spectral density after long times. As detailed in the Appendix~\ref{appendixd5},  pure classical noise does not generate entanglement, hence we examine the case of low temperature where both classical and quantum noise are present. With the spatially correlated Markovian noise, the coherent interaction between the two qubits becomes time-independent, and is described by the Hamiltonian 
\(H_{xy}=\mathcal{J} \hat{\sigma}_1^+ \hat{\sigma}_2^-+\mathcal{J}^*\hat{\sigma}_1^-\hat{\sigma}_2^+.\)
  The coupling strength 
\(\mathcal{J} =\frac{\mathcal{G}^R_{12}(\Omega) +\mathcal{G}^R_{21}(-\Omega)   }{2\hbar^2} \  \) 
 is complex-valued in general and denoted as $\mathcal{J}\equiv \mathcal{J}_s+i\mathcal{D}$.  The contributions $\mathcal{J}_s$ and Dzyaloshinskii–Moriya (DM)  interaction  $\mathcal{D}$ are symmetric and antisymmetric with respect to exchange of the two qubits, respectively,
\(  H_{xy} = \mathcal{J}_s ( \hat{\sigma}^x_1 \hat{\sigma}^x_2 +\hat{\sigma}^y_1 \hat{\sigma}^y_2   ) + \mathcal{D} \hat{z} \cdot \vec{\hat{\sigma}}_1\times \vec{\hat{\sigma}}_2,   \)
 where the DM interaction only arises when the inversion symmetry of the environment is broken.

The decay and absorption rates become time-independent when noise is assumed to be Markovian (or when the evolution time is long for generic noise, $t\gg 1/\Omega$) and are given by Eq.~\eqref{eq:markovianrate}. 
At temperatures lower than the Rabi frequency, $k_BT\leq \hbar \Omega$, (for instance, when $\Omega/2\pi\sim 2\,\text{ GHz}$ and the temperature is below 100 mK), the rates are further reduced to $\gamma_{ij}^\downarrow\approx 2 S^Q_{ij}(\Omega)$ and $\gamma_{ij}^\uparrow \approx 0$.
In this situation, the superoperators are given by the standard Lindbladians~\cite{Heinz}:
\(   \al{ \mathcal{L}_{ij}\rho &= \gamma^\downarrow \sum_{i}\Big[ \hat{\sigma}_i^- \rho \hat{\sigma}_i^+ -\frac{1}{2}\{  \hat{\sigma}_i^+\hat{\sigma}_i^-,\rho    \}   \Big]   \\
   &\; +  \gamma_{12}^\downarrow \sum_{i\neq j}\Big[ \hat{\sigma}_j^- \rho \hat{\sigma}_i^+ -\frac{1}{2}\{  \hat{\sigma}_i^+\hat{\sigma}_j^-,\rho    \}   \Big] ,}     \)
where the local decay rate $\gamma^\downarrow\equiv \gamma^\downarrow_{ii}$ and the collective decay rate $\gamma^\downarrow_{12}>0$ are determined by the local and spatially correlated noise, respectively. We have absorbed the phase of $\gamma^\downarrow_{12}$ into the definition of $\sigma^{\pm}_{i}$. The completely positive evolution dictates that the correlated decay is weaker than the local decay $\gamma^\downarrow_{12}\leq \gamma^\downarrow$, which is also guaranteed by the thermodynamic stability of the environment~\cite{zou2022prb}.

  Similar to the coherent interaction, it is convenient to  symmetrize and antisymmetrize the superoperators, yielding
   \(  \mathcal{L}\rho=\sum_{\alpha\in\{S, A\}} \Gamma_{\alpha} \Big[  \hat{\sigma}^-_\alpha \rho \hat{\sigma}^+_\alpha - \frac{1}{2} \{ \hat{\sigma}_\alpha^+ \hat{\sigma}_\alpha^-, \rho   \}    \Big] ,    \)
where $ \sigma^+_{S, A}=(\sigma_1^+\pm \sigma_2^+) /\sqrt{2}$ and $\Gamma_{S, A}= \gamma^\downarrow \pm \gamma^\downarrow_{12}$. We note that the triplet state $\ket{T}\equiv ( \ket{\uparrow\downarrow} +\ket{\downarrow\uparrow} )/\sqrt{2}$ is superradiant decaying at rate $\Gamma_S$, while the singlet state, $\ket{S}\equiv ( \ket{\uparrow\downarrow} -\ket{\downarrow\uparrow} )/\sqrt{2}$ is subradiant decaying at rate $\Gamma_A$. We also remark that these two states are decoupled from each other in $\mathcal{L}\rho$, and  are also eigenstates of the symmetric interaction characterized by $\mathcal{J}_s$. 
However, the DM interaction, which is parity-odd, exchanges these two states $(\hat{z}\cdot   \vec{\hat{\sigma}}_1\times \vec{\hat{\sigma}}_2) \ket{T, S}\propto \ket{S, T}$.

In the following subsection, we analytically study the interplay between the symmetric interaction, the DM interaction, and the local and correlated decay processes.    For the sake of concreteness, we assume the initial state is a trivial product state $\ket{\uparrow\downarrow}$. We denote  the density matrix as $\rho=G_t\ket{T}\bra{T}+G_s\ket{S}\bra{S}+\big(G_{ts}\ket{T}\bra{S}+\text{H.c.} \big)+\Delta\rho$, where $\Delta \rho$ stands for other elements of the density matrix.  In this scenario,  the concurrence of the two-qubit system can be shown to take the simple form of $\mathcal{\mathcal{C}}[\rho]=|\mathcal{C}_R+i \mathcal{C}_I|$, with 
\( \mathcal{C}_R\equiv G_{s}-G_t, \;\;\;\text{and}\;\;\; \mathcal{C}_I\equiv 2\Im G_{ts}, \label{eq28}\)
which indicates that the entanglement of the qubits has two independent contributions (see Appendix~\ref{appendixd1} for detailed derivations). To gain some physical understanding of this expression, we observe that the first contribution arises from the asymmetry in the populations of the triplet state $\ket{T}$ and singlet state $\ket{S}$, which is proportional to $\ket{\uparrow\downarrow}\bra{\downarrow\uparrow}+\ket{\downarrow\uparrow}\bra{\uparrow\downarrow}$. The second contribution comes from a finite imaginary part of $G_{ts}$, which is proportional to $\ket{\uparrow\downarrow}\bra{\downarrow\uparrow}-\ket{\downarrow\uparrow}\bra{\uparrow\downarrow}$. Both indeed characterize the coherence (superposition) of $\ket{\uparrow\downarrow}$ and $\ket{\downarrow\uparrow}$.

\begin{figure*}
	\centering\includegraphics[width=0.92\linewidth]{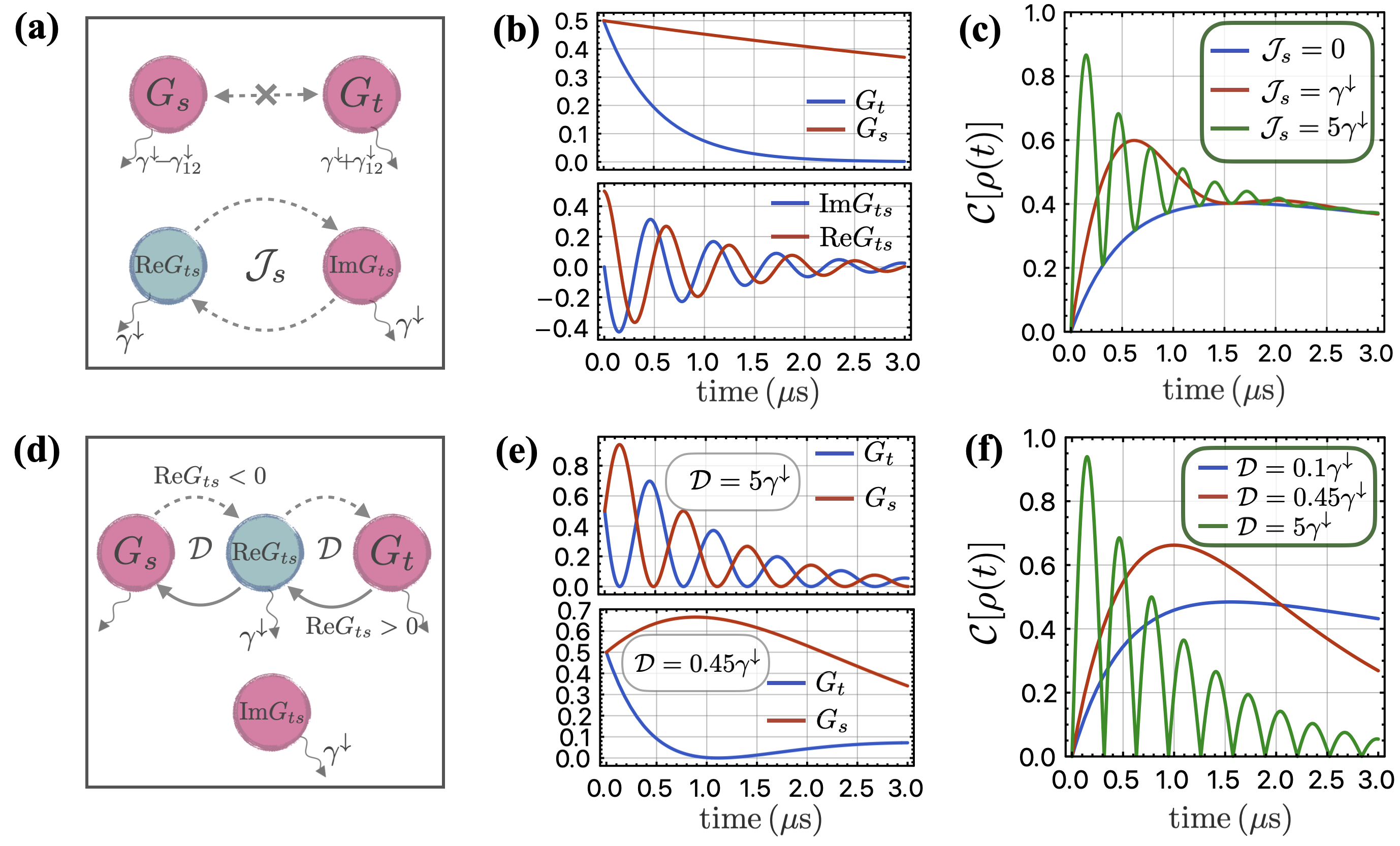}
 \caption{Entanglement dynamics of resonantly driven qubits with initial state $\ket{\uparrow\downarrow}$. (a)-(c) Qubit dynamics in the absence of DM interaction $\mathcal{D}$. (a) Schematic for the coupled dynamics of relevant density matrix elements. In the absence of the DM interaction $\mathcal{D}=0$, the dynamics of $G_s$ and $G_t$ are decoupled from each other, decaying independently at rates of $\gamma^\downarrow-\gamma^\downarrow_{12}$ and  $\gamma^\downarrow+\gamma^\downarrow_{12}$, respectively. The real and imaginary parts of $G_{ts}$ are coupled to each other through the symmetric coupling $\mathcal{J}_s$, while they decay at the same rate $\gamma^\downarrow$.   (b) The upper panel illustrates the decay of the superradiant state $\ket{T}$ and the subradiant state $\ket{S}$, which is independent of the parameter $\mathcal{J}_s$. The lower panel shows the oscillations of the real and imaginary parts of $G_{ts}$ with frequency $2\mathcal{J}_s$, where we use parameter $\mathcal{J}_s=5 \gamma^\downarrow$. (c) Entanglement quantified by the concurrence $\mathcal{C}[\rho(t)]$ between two qubits as a function of time with varying parameter $\mathcal{J}_s$. The oscillation frequency of the entanglement is $\propto \mathcal{J}_s$. The entanglement is bounded below by $|G_s-G_t|$ at any time. At large time $t\gg 1/\gamma^{\downarrow}$, the oscillation is insignificant and the dynamics is dominated by the local and correlated noise. 
  (d)-(f) Qubit dynamics in the absence of symmetric interaction. (d) The dynamics of $G_t$ and $G_s$ are coupled to each other via $\Re G_{ts}$, in the presence of the DM interaction, while $\Im G_{ts}$ is decoupled from other elements. Assuming $\mathcal{D}>0$ without loss of generality, the probability in $\ket{S}$ would flow to $G_t$ when $\Re G_{ts}<0$, whereas the probability flows in the opposite direction when $\Re G_{ts}>0$.  The changing rate of $\Re G_{ts}$ is determined by the difference between $G_t$ and $G_s$, $\Re \partial_t{G}_{ts}=-\gamma^\downarrow \Re G_{ts} +\mathcal{D} (G_t-G_s)$. (e) The upper panel shows the oscillations of $G_t$ and $G_s$ when $\mathcal{D}=5\gamma^\downarrow$.  The lower panel shows the their time evolution when the DM coupling is small $\mathcal{D}=0.45\gamma^\downarrow$, where we do not see the oscillation behavior as the dynamics is overdamped. (f) Entanglement between two qubits as a function of time with varying strength of DM coupling $\mathcal{D}$. The green curve shows the oscillation of entanglement, where the DM coupling is large and the dynamics is underdamped.The red curve is the critical point, where the entanglement stops oscillating and behaves as $\propto t e^{-\gamma^\downarrow t}$.  The blue curve is when the DM coupling is small, where the dynamics is overdamped.   Parameters used in all figures: $\gamma^\downarrow=1\,\mu \text{s}^{-1}$ and $\gamma^\downarrow_{12}=0.9 \gamma^\downarrow$.  }
  \label{fig4}
\end{figure*}

\subsection{The role of symmetric exchange interaction}
In this subsection, we analyze one basic scenario in which the environment possesses inversion symmetry, resulting in a real correlated noise spectral density $S_{12}(\omega)$. In this case, the DM interaction is absent.   The dynamics of the singlet state $G_s$ and triplet state $G_t$ are decoupled due to the symmetry, while the real and imaginary parts of $G_{ts}$ are coupled to each other due to the symmetric exchange coupling $\mathcal{J}_s$, as illustrated in Fig.~\ref{fig4} (a). The complete dynamics is analytically solved in the Appendix~\ref{appendixd2}.

The impact of correlated quantum noise on the two-qubit dynamics is twofold. Firstly, the noise enters through the collective decay $\gamma^\downarrow_{12}$, which results in different decay rates for the population of the singlet and triplet states, as demonstrated in the upper panel of Fig. \ref{fig4} (b), leading  to an increase in entanglement.
 Secondly, the noise enters through the coupling $\mathcal{J}_s$, which gives rise to oscillation of $\Im G_{ts}$, causing the entanglement to also oscillate. This is illustrated in the lower panel of Fig. \ref{fig4} (b). To quantitatively analyze how the quantum coherence in the system evolves, we deduce the equation of motion for $\mathcal{C}_R$ and $\mathcal{C}_I$, which take the following simple form:
\( \al{ \ddot{ \mathcal{C} }_R+2\gamma^\downarrow \dot{\mathcal{C}}_R +(\gamma^{\downarrow 2} -\gamma^{\downarrow 2}_{12}) \mathcal{C}_R=0, \\  
         \ddot{ \mathcal{C} }_I+2\gamma^\downarrow \dot{\mathcal{C}}_I +(\gamma^{\downarrow 2} +4 \mathcal{J}_s^2  ) \mathcal{C}_I=0 , } \)
 representing an overdamped and an underdamped harmonic oscillators, respectively.
The local noise $\gamma^\downarrow$ serves as a ``friction" force,  impeding the increase of $\mathcal{C}_R$ and $\mathcal{C}_I$ and, thus, the entanglement.  Surprisingly, the correlated quantum noise ($\gamma^\downarrow_{12}$ and $\mathcal{J}_s$) acts as an active force that facilitates the increase of entanglement.
Considering the initial state $\ket{\uparrow\downarrow}$, these equations can be solved and reveal the following  entanglement dynamics encoded in the concurrence:
\( \mathcal{C}[\rho(t)]=e^{-\gamma^\downarrow t} \sqrt{ \sinh^2(\gamma^\downarrow_{12}t)  + \sin^2(2\mathcal{J}_st)  }.  \label{eq30} \)
We observe that while the local noise has a detrimental effect on entanglement, the correlated quantum  noise-induced correlated decay and coherent coupling both have a beneficial effect. Interestingly,  in the absence of coherent coupling $\mathcal{J}_s=0$, a sizable amount of entanglement can be generated in the pure dissipative evolution due to the correlated decay, corresponding to the blue curve in Fig.~\ref{fig4} (c). It also provides a lower bound on the entanglement when $\mathcal{J}_s$ is finite, as evidenced by the green and red curves in Fig.~\ref{fig4}(c), which display entanglement oscillations with a frequency proportional to $\mathcal{J}_s$. We  remark that, at the long-time evolution limit, the entanglement scales as $\mathcal{C}[\rho(t\rightarrow \infty)]\propto \text{exp}[-(\gamma^\downarrow -\gamma^\downarrow_{12} )t]$.  This persistence can be attributed to  the slow decay of the singlet state, which is long-lived  when the correlated noise is comparable to the local noise, $\gamma^\downarrow_{12}\rightarrow \gamma^\downarrow$.

\begin{figure}
	\centering\includegraphics[width=0.8\linewidth]{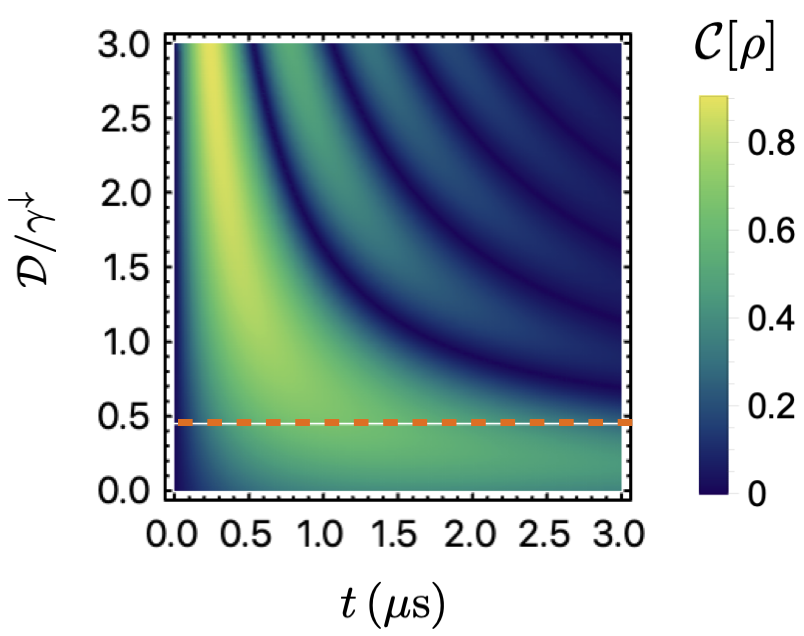}
 \caption{ Entanglement dynamics of two driven qubits under DM interaction with initial state $\ket{\uparrow\downarrow}$. The plot shows the time evolution of entanglement as a function of DM interaction strength $\mathcal{D}$. The dashed orange line at $\mathcal{D} = 0.45\gamma^\downarrow$ separates the underdamped and overdamped regimes. When $\mathcal{D}$ is above the dashed line, the entanglement exhibits oscillations with increasing frequency as $\mathcal{D}$ increases. In contrast, below the dashed line, the entanglement decays on a longer timescale and does not exhibit oscillations. Parameters we use in the figure are $\gamma^\downarrow=1\, \mu \text{s}^{-1} $ and $\gamma^\downarrow_{12}=0.9\gamma^\downarrow$.  }
  \label{fig5}
\end{figure}

\subsection{The role of Dzyaloshinskii–Moriya interaction}
In this subsection, we study another basic scenario where the DM interaction is present while the symmetric interaction vanishes $\mathcal{J}_s=0$. The dynamics of the population of the singlet state $G_s$ and of the triplet state $G_t$ are now coupled to each other due to parity-breaking interaction $\mathcal{D}$. 
The probability flows from the triplet state to singlet state when $\Re G_{ts}>0$, and in the opposite direction when $\Re G_{ts}<0$, while the dynamics of $\Re G_{ts}$ is governed by the relative population of these states and can be expressed as $\Re \partial_t{G}_{ts}=-\gamma^\downarrow \Re G_{ts} +\mathcal{D} (G_t-G_s)$, as shown in Fig.~\ref{fig4} (d). 
 The element $\Im G_{ts}$ is decoupled from other elements and remains zero when the initial state is $\ket{\uparrow\downarrow}$. We solve the coupled dynamics analytically in Appendix~\ref{appendixd3}.

The two-qubit dynamics is affected in two ways by the presence of correlated quantum noise. Firstly, similar to the pure symmetric exchange case, the correlated noise-induced collective decay $\gamma^\downarrow_{12}$ gives rise to different decay rates of $G_t$ and $G_s$, which can lead to the generation of entanglement. On the other hand,  the correlated noise-induced DM interaction $\mathcal{D}$ causes an oscillation between the singlet and triplet state, which can interfere with the effect of $\gamma^\downarrow_{12}$ in a nontrivial manner. Remarkbaly, one can construct a rather simple equation of motion for $\mathcal{C}_R$ from the coupled complex dynamics to quantify the entanglement evolution:
\( \ddot{ \mathcal{C} }_R+2\gamma^\downarrow \dot{\mathcal{C}}_R +(\gamma^{\downarrow 2} -\gamma^{\downarrow 2}_{12} +4\mathcal{D}^2 ) \mathcal{C}_R=0.  \label{eq31} \)
We observe that the quantum correlated noise-induced collective decay and DM interaction compete with each other, while the local noise $\gamma^\downarrow$ still acts as a ``friction" force as before. When the coherent interaction dominates $\mathcal{D}>\gamma^\downarrow_{12}/2$, the system oscillates between the singlet and triplet states, resembling an underdamped oscillator, as illustrated in the upper panel of Fig.~\ref{fig4} (e). In the regime where $\mathcal{D}<\gamma^\downarrow_{12}/2$, the system can be described by an overdamped oscillator. Whenever the probability flows to the triplet state, it quickly decays due to the strong dissipation $\gamma^\downarrow+\gamma^\downarrow_{12}$, preventing the probability from returning to the singlet state, and therefore, it does not exhibit any oscillatory behavior, as illustrated in the lower panel of Fig.~\ref{fig4} (e).  With the initial state $\ket{\uparrow\downarrow}$, we solve the entanglement dynamics, which reveals three distinct dynamic regimes: 
\( \mathcal{C}[\rho(t)]\!=\! e^{-\gamma^\downarrow t} \times \begin{cases}   | ( \gamma^\downarrow_{12} +2\mathcal{D} ) \sin\omega_r t  |/\omega_r,\! \!\!\!\! \!\!&2 |\mathcal{D}|>\gamma^\downarrow_{12},  \\
   2\gamma^\downarrow_{12} t ,\!  &  2 |\mathcal{D}|=\gamma^\downarrow_{12},  \\
    | ( \gamma^\downarrow_{12} +2\mathcal{D} )| \sinh\kappa t  /\kappa,  &2 |\mathcal{D}|<\gamma^\downarrow_{12} ,  \end{cases}   \)
with $\omega_r=\sqrt{4\mathcal{D}^2-\gamma^{\downarrow 2}_{12}}$ and $\kappa=\sqrt{ \gamma^{\downarrow 2}_{12} - 4\mathcal{D}^2  } $. In the underdamped regime, the entanglement exhibits oscillations at a frequency of $\omega_r$, as illustrated by the green curve in Fig.~\ref{fig4} (f), where the odd and even peaks correspond to the singlet and triplet states, respectively. The entanglement decays on a characteristic timescale of $1/\gamma^\downarrow$. At the critical point where $2|\mathcal{D}|=\gamma^\downarrow_{12}$, the oscillation behavior ceases, and the entanglement follows a scaling of $\propto te^{-\gamma^\downarrow t}$, as depicted by the red curve in Fig.~\ref{fig4} (f). The loss of oscillation can be clearly observed in Fig.~\ref{fig5}, where the dashed orange line represents the critical point; entanglement exhibits oscillations above this point, while there is no oscillation below it.
In the overdamped regime, the entanglement exhibits a scaling behavior of $\propto \text{exp}[-(\gamma^\downarrow-\kappa)t]$, with an extended lifetime of $1/(\gamma^\downarrow-\kappa)$, while the maximal entanglement it can reach is reduced in the dynamics, as shown by the blue curve in Fig.~\ref{fig4} (f).

Our findings suggest that despite the issues that  noise can produce in quantum information processing, if properly harnessed, correlated quantum noise provides a useful resource to generate a significant long-lived entanglement, opening up to a several opportunities for optimizing quantum computation. We stress that the process of generating entanglement can be controlled on-demand by switching the driving on and off.
 Compared to the pure-dephasing dynamics, where the correlated dephasing cannot build up coherence despite being rooted in the correlated quantum noise, in this section we find that, interestingly, the correlated decay induced by the transverse noise can generate sizable entanglement by causing the singlet and triplet states to decay at different rates. In particular, while both rooted in correlated quantum noise and beneficial for entanglement generation, the competition between the correlated decay and coherent coupling leads to distinct dynamical regimes. 
From the two fundamental cases presented here,  one can extrapolate to dynamics when both symmetric exchange and DM interaction are present, and we leave the detailed discussion of this situation to Appendix~\ref{appendixd4}.

\section{  Driven spin qubits subject to correlated $1/f$ noise}~\label{secv}
Building upon the insights obtained from the previous study on Markovian noise, in this section, we  investigate the impact of spatially and temporally correlated $1/f$ noise, which is non-Markovian in nature.  We assume that the two driven qubits are located within a few hundred nanometers of each other, and thus the spatially correlated noise is comparable to the local noise $S_{12} \approx S_{ii}$ in the relevant frequency range. We present an analytical investigation into two situations, in the first one, only classical $1/f$ noise is present, while in the second one, we  include also  quantum $1/f$ noise, comparable to the classical noise. This situation can occur at low temperatures, when $k_B T\leq \hbar\Omega$, where the ability of qubits to emit and absorb energy is different as discussed in Sec.~\ref{Sec_2.1}.  For simplicity, we  assume that the spectral density of the correlated noise is real, and we focus on its temporal correlations. 
The detailed derivations of the results in this section are sketched in Appendix~\ref{appendixe}.

\begin{figure}
	\centering\includegraphics[width=\linewidth]{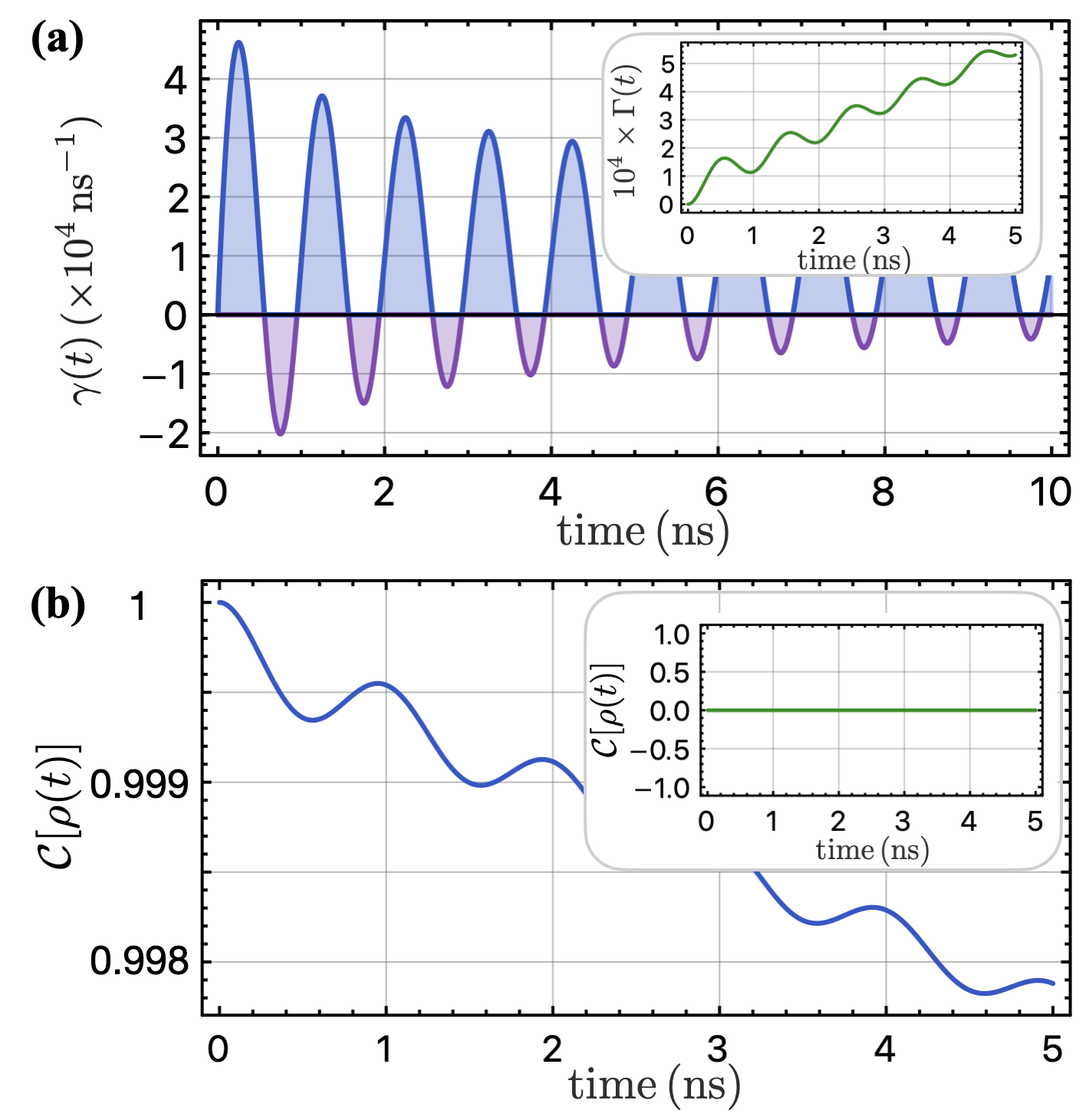}
 \caption{ Entanglement dynamics of two driven qubits under pure classical $1/f$ noise. (a)  Temporal evolution of the decay rate $\gamma(t)$ depicted as a function of time, where the occurrence of negative values is identified within the marked purple intervals. However, the integral of the decay rate $\Gamma(t)$, as highlighted in the inset of (a), must remain nonnegative to ensure the complete positivity of the dynamics.  (b) Entanglement between two qubits as a function of time with the initial state being a Bell state $\ket{\psi_0}=(\ket{\uparrow\downarrow}+i\ket{\downarrow\uparrow})/\sqrt{2}$, and the inset shows the entanglement with initial state $\ket{\uparrow\downarrow}$. Parameters that are used in the plots: $\hbar/\sigma=100\,\text{ns}$, $\omega_l/2\pi=1\,\text{MHz}$, and $\Omega/2\pi=1\,\text{GHz}$.}
  \label{fig6}
\end{figure}

\subsection{Classical $1/f$ noise}\label{secv1}
In the presence of purely classical $1/f$ noise, the two-qubit dynamics is governed by the TCL master equation~\eqref{eq:mastereq} with vanishing coherent coupling $\mathcal{J}(t)=0$ and equal local and correlated absorption and decay rates, denoted as $\gamma(t)\equiv \gamma^\downarrow_{ij} =\gamma^\uparrow_{ij}$, with following form: 
\(\al{  \gamma(t)&=\frac{2}{\hbar^2} \int^\infty_0 \frac{d\omega}{2\pi} S^C_{ii}(\omega) \big[  F_s(\omega-\Omega,t) +  F_s(\omega+\Omega,t)   \big]  \\
                           &= \frac{4\sigma^2}{\hbar^2 \Omega} \big[  \text{Si}(\Omega t) -\sin(\Omega t) \text{Ci}(\omega_l t)   \big].  } \label{eq33}  \)
Here, $\text{Si}(x)\equiv \int^x_0 d\tau \sin(\tau)/\tau$ is the sine integral function.  
One surprising feature of the rate $\gamma(t)$ is that it can take negative values  during finite time intervals, denoted by the purple regions in Fig.~\ref{fig6} (a). This behavior is an indication of non-Markovian memory effects, reflecting the exchange of information between the two qubits and the environment~\cite{piilo2008non}. Nevertheless, the time integral of $\gamma(t)$, denoted as \( \Gamma(t)\equiv \int^t_0ds\, \gamma(s),\)
 must remain non-negative due to the complete positivity requirement of the system dynamics~\cite{PhysRevA.57.120,bylicka2014non}. This is illustrated in the inset of Fig.~\ref{fig6} (a). 
 
Based on the insights gained from the previous sections, we anticipate that entanglement generation would not be possible with purely classical correlated noise. This is indeed the case even when the noise is temporally correlated, as shown in the inset of Fig.~\ref{fig6} (b), where the entanglement remains to be zero with a trivial initial product state $\ket{\uparrow\downarrow}$. A new feature arising from the non-Markovian nature is the occurrence of oscillations in the decay of entanglement,  when the initial state is entangled. For concreteness, we   initialize the system  to be the Bell state $\ket{\psi_0}=(\ket{\uparrow\downarrow}+i\ket{\downarrow\uparrow})/\sqrt{2}$. The dynamics of the  entanglement is then given by
\(  \mathcal{C}[\rho(t)]= \frac{1}{3} \Big[\sqrt{  4 e^{-6\Gamma(t)} \sinh^2 (3\Gamma)  +9 e^{-4\Gamma(t)}  } + e^{-6\Gamma(t)}-1  \Big] , \label{eq35} \)
which is depicted in Fig.~\ref{fig6} (b). To gain some insights, let us consider a specific quantum trajectory.  During time intervals where the rate $\gamma(t)>0$, the system can undergo quantum jumps, which can lead to the loss of coherence and a transition from an entangled state such as $\ket{\psi_0}$ to a trivial state like $\ket{\downarrow\downarrow}$ with a finite probability. Conversely, during the negative rate $\gamma(t)<0$ later on, the quantum jump process can be interpreted as a jump in the reverse direction~\cite{daley2014quantum,piilo2008non,piilo2009open,milz2021quantum}, $\ket{\psi_0}\leftarrow \ket{\downarrow\downarrow}$, with a finite probability of restoring the lost superposition due to the non-Markovian memory effect.  As a result, the entanglement exhibits a temporary increase during the decoherence process, as illustrated in Fig.~\ref{fig6} (b).

\begin{figure*}
	\centering\includegraphics[width=0.98\linewidth]{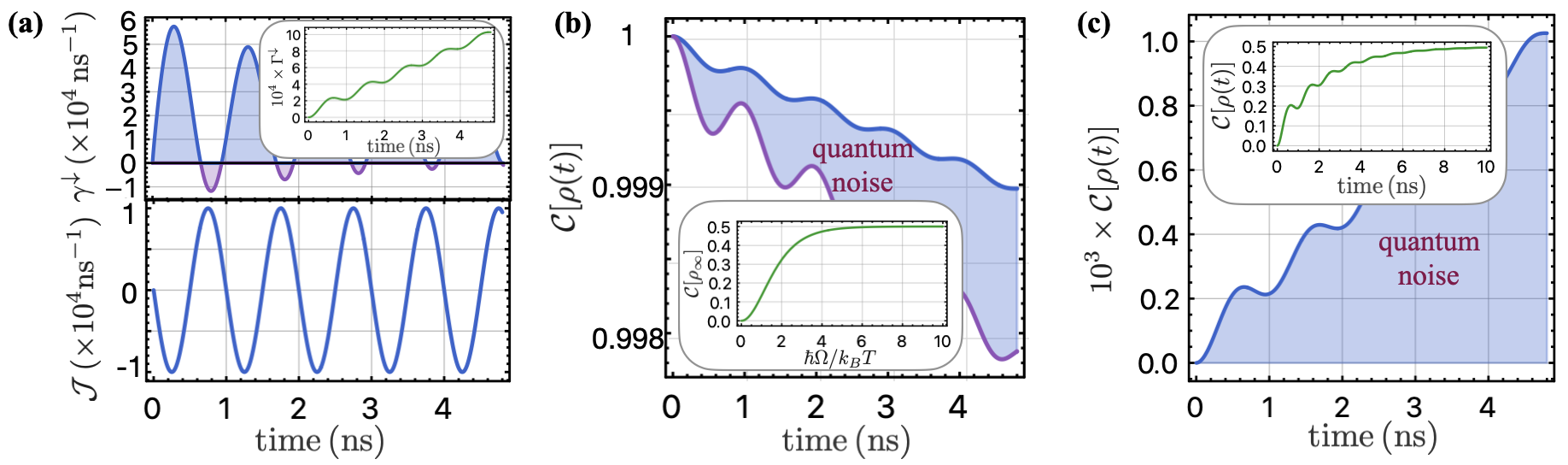}
 \caption{Analysis of driven qubit dynamics in the presence of quantum noise. (a) The upper panel depicts the time-dependent decay rate $\gamma^\downarrow$, which can take negative values for certain time intervals (indicated by purple shading). Its time integral $\Gamma^\downarrow$ is positive at all times to ensure the complete positivity of the dynamics. The lower panel shows the coherent coupling $\mathcal{J}$ between two qubits as a function of time. (b) Time-dependent entanglement for a maximally entangled initial state $\ket{\psi_0}=(\ket{\uparrow\downarrow}+i\ket{\downarrow\uparrow})/\sqrt{2}$. The blue shaded area demonstrates the impact of quantum noise, with the blue and purple curves representing cases with and without quantum noise, respectively. The inset displays the final steady-state entanglement as a function of temperature, which is zero in the classical limit and 1/2 in the quantum regime.   (c) Entanglement evolution as a function of time for an initial state $\ket{\uparrow\downarrow}$. The effect of quantum noise is illustrated by the blue shaded region (the entanglement remains at zero when only classical noise is present). The inset reveals a final entanglement value of 1/2 with $\hbar/\sigma=3\, \text{ns}$ within a time $\sim 10\, \text{ns}$. The plotted results are obtained using the following parameters: $\hbar/\sigma=100\,\text{ns}$, $\omega_l/2\pi=1\,\text{MHz}$, and $\Omega/2\pi=1\,\text{GHz}$. }
  \label{fig7}
\end{figure*}

\subsection{Quantum $1/f$ noise}\label{secv2}
We now investigate the impact of quantum correlated noise in the quantum regime where $S^Q\approx S^C$. The dynamics of the two-qubit system is governed by the same master equation~\eqref{eq:mastereq}, with the decay and absorption rates defined as
\( \al{ \gamma_{ij}^\downarrow(t)=\frac{4}{\hbar^2}\int^\infty_0 \frac{d\omega}{2\pi} S^Q_{ij}(\omega) F_s(\omega-\Omega, t), \\
\gamma_{ij}^\uparrow(t)=\frac{4}{\hbar^2}\int^\infty_0 \frac{d\omega}{2\pi} S^Q_{ij}(\omega) F_s(\omega+\Omega, t). }\)
Considering that the filter function $F_s(\omega+\Omega, t)$ is peaked at $\omega=-\Omega$ which lies outside the range of integration, we approximate the absorption rate $\gamma^\uparrow_{ij}$ to be zero to enable complete analytical solution of the dynamics. In the strong correlated noise regime considered in this section, we assume that the local and correlated decay rates are equal and we denote them as $\gamma^\downarrow(t)\equiv \gamma^\downarrow_{ij}(t)$, which is evaluated to be:
\(  \gamma^\downarrow(t)= \frac{4\sigma^2}{\hbar^2\Omega} \big[ \pi(1-\cos\Omega t)/2 -\sin \Omega t \text{Ci}(\omega_l t) +\text{Si}(\Omega t)   \big].   \label{eq37}\)
The upper panel of Fig.~\ref{fig7} (a) shows that the rate $\gamma^\downarrow(t)$ can be negative temporarily, while its time integral defined as
\(\Gamma^\downarrow(t)\equiv \int^t_0 ds\, \gamma^\downarrow(s),\)
must be positive due to the complete positivity of the dynamics, as illustrated in the inset. 
In the presence of quantum correlated noise, the coherent coupling $\mathcal{J}(t)$  in the Hamiltonian~\eqref{eq:hamiltonian} is nonvanishing and  given by:
\( \al{ \mathcal{J}(t) & = \frac{2}{\hbar^2}\int^\infty_0 \frac{d\omega}{2\pi} S^Q(\omega) \big[  F_c(\omega-\Omega, t) +  F_c(\omega+\Omega, t)   \big]  \\
                                &= -\frac{2\pi\sigma^2}{\hbar^2 \Omega} \sin\Omega t ,} \label{eq39}  \)
  which oscillates with a frequency of $\Omega$ and takes values comparable to the decay rate as shown in the lower panel of Fig.~\ref{fig7} (a). To demonstrate the effect of quantum noise, we first initialize the system into the Bell state $\ket{\psi_0}=(\ket{\uparrow\downarrow}+i\ket{\downarrow\uparrow})/\sqrt{2}$ and investigate how the entanglement decays in the presence of both classical and quantum $1/f$ noise.  We show the entanglement in this case is given by 
       \(  \mathcal{C}[\rho(t)]=e^{-\Gamma^\downarrow(t)} \sqrt{  \sinh^2\Gamma^\downarrow(t) +\cos^2\Phi(t)   },   \label{eq40}  \)
  with  the phase $\Phi(t)$ defined as
 $ \Phi(t)\equiv \int_0^tds\, \mathcal{J}(s)=    2\pi \sigma^2 (\cos\Omega t -1)/\hbar^2\Omega^2.  $
It is shown as the blue curve in Fig.~\ref{fig7} (b), with the purely classical $1/f$ noise case also shown for contrast as the purple curve.
The temporary increase in the entanglement during the decoherence process is also observed in the presence of quantum correlated noise. In the case of strong correlated quantum $1/f$ noise, decoherence occurs at a much slower rate, and the net effect of quantum noise is reflected in the shaded blue region in Fig.~\ref{fig7} (b). The entanglement is long-lived, and the final entanglement approaches 1/2 due to the long-lived singlet state with a decay rate of $\gamma^\downarrow_{ii}-\gamma^\downarrow_{12}$, which is almost zero when the correlated noise is comparable to the local noise. For arbitrary temperature,  the residual entanglement is shown to be: 
                                \( \al{ \mathcal{C}[\rho_\infty]&=\frac{1}{2}-\frac{3}{4\cosh(\beta \hbar \Omega) +2  }   \\
                                                                                &=  \frac{2[S^Q(\Omega)]^2}{3[S^C(\Omega)]^2+[S^Q(\Omega)]^2},   }  \label{eq41}  \)   
                                which is shown in the inset of Fig.~\ref{fig7} (b). Here we have invoked the fluctuation-dissipation theorem $S^C_{ij}(\omega)=\coth(\beta\hbar \omega/2)S^Q_{ij}(\omega)$.   It can be clearly observed that in the classical limit where $k_B T \gg \hbar \Omega$, $S^Q=0$, the final entanglement is zero as expected. Furthermore, we also conclude  that there is always a finite, long-lasting entanglement present when the correlated quantum noise (comparable to local quantum noise) is finite.
 
To investigate the entanglement generation by spatially correlated quantum $1/f$ noise, we consider a simple initial state $\ket{\uparrow\downarrow}$. The entanglement then is given by
                                \(  \mathcal{C}[\rho(t)]=e^{-\Gamma^\downarrow(t)} \sqrt{  \sinh^2\Gamma^\downarrow(t) +\sin^2\Phi(t)   }.   \label{eq42} \)
We note that there are temporary decrease in the growing entanglement, which finally reaches $1/2$, as shown in Fig.~\ref{fig7} (c) and its inset. This can also be attributed to the non-Markovian memory effect. Considering a quantum trajectory, when the decay rate is positive $\gamma^\downarrow(t)>0$, the system can undergo quantum jumps, taking the product state to an entangled state in the presence of correlated quantum noise. Later, when the decay rate is negative $\gamma^\downarrow(t)<0$, the state jumps back to a product state from the entangled state, which leads to the temporary dip in the entanglement growth. Our findings demonstrate the potential of utilizing the correlated quantum $1/f$ noise, which is ubiquitous in solid-state quantum computing platforms,  to generate significant entanglement or delay the decoherence process in two-qubit systems. We also illustrate the effects of the non-Markovianity of the noise in the dynamics of two driven qubits.

\section{Conclusion}\label{secvi}
In this paper, we have presented a comprehensive analytical study of the two-qubit dynamics subject to both local and non-local spatially correlated noise. Our analysis is based on a time-local TCL master equation that is applicable for generic noise spectra, including both Markovian and non-Markovian noise. We explored how the classical and quantum correlations stored in the noise dictate the qubit dynamics. Our results reveal that, at high temperatures compared to the relevant qubit energy when only classical correlations are present, the correlated noise only modifies the decoherence rate without leading to any entanglement between qubits. 
One can therefore operate the qubits at warmer temperatures to effectively suppress undesired crosstalk between qubits caused by correlated noise~\cite{undseth2023hotter}.

In the case of low temperatures, when both classical and quantum correlations are present in the noise, the  correlated quantum noise introduces various new effects. These include the coherent \textit{Ising interaction} and \textit{correlated dephasing} in the case of purely dephasing noise, as well as the coherent \textit{symmetric exchange}, \textit{DM interaction}, and \textit{correlated relaxation} in the case of transverse noise. We show that these dissipative interactions can be turned on and off on-demand by resonantly driving the qubits. We have illustrated the effects of {these interaction} by solving the two-qubit dynamics analytically. Specifically, our analysis has demonstrated that only the {noise-induced} Ising interaction, not the correlated dephasing, can lead to finite entanglement generation. However, for the case of transverse noise, we found that both coherent interactions and correlated relaxation give rise to significant long-lived entanglement. Their competitions generally lead to different dynamical regimes of the two-qubit system. 
Therefore, by driving the qubits and operating the system at lower temperatures, one can exploit the quantum correlations stored in the noise for various quantum information applications.
We finally also studied the non-Markovian memory effect by investigating the correlated $1/f$ noise, where the decay rate can be negative temporarily. 

Our work provides a comprehensive understanding of how the classical and quantum correlated noise affects the  qubit dynamics.   Our analysis enables  the development of effective strategies for utilizing the noise correlations in quantum information processing or for mitigating their potentially harmful effects, paving the way towards the design of robust and scalable quantum technologies. 

Future work will explore the effect of long-ranged noise, beyond nearest-neighbor, in multiple qubit systems. This case is critical for future experiments aiming to scale up quantum processors, especially because recent measurements highlighted the presence of long-range noise in spin qubits in quantum dots~\cite{rojas2023spatial}. Investigating how the correlated noise affects standard quantum operations, such as the fidelity of two-qubit gates is therefore critical~\cite{PhysRevA.57.120}. Additionally, it is important to understand how the correlated noise interferes with existing strategies developed to suppress the impact of single-qubit noise, such as quantum error correction codes, dynamical decoupling, {and sweet spots}. Our work provides a solid foundation for future research in these directions and will serve as a starting point to address these challenges {towards large-scale quantum computers.}

\begin{acknowledgements}
This work was supported by the Georg H. Endress Foundation and by the Swiss National Science Foundation, NCCR SPIN (grant number 51NF40-180604).
\end{acknowledgements}

\appendix
\section{Spatially correlated noise in different dimensions}
\label{appendix1}
In Sec.~\ref{Sec_2.1} in the main text, we introduce the local and spatially correlated  noise spectral densities $S_{ij}(\omega)$, defined in Eq.~\eqref{Eq:9}. To be specific, we present the relation between the local and correlated noise in two dimensional architectures in the main text.  In this section, we provide a detailed illustration of their relations in various dimensions, assuming a linear spectrum of the environment as $\omega_{\vb k}=c_s|\vb k|$. This allows us to write the noise spectral density as:
\(\al{ S_{ij}(\omega)=&2\pi |g_k|^2 [ n_B(\omega)+1] \sum_{\vb k} e^{-i\vb k\cdot (\vb r_j-\vb r_i)} \delta(\omega-\omega_k) \\
                               &\; +2\pi |g_{-k}|^2 n_B(-\omega)   \sum_{\vb k} e^{i\vb k\cdot (\vb r_j-\vb r_i)} \delta(\omega+\omega_k),   }\)
where $k=\omega/c_s$ and  we have assumed the coupling $g_{\vb k}$ only depends on the magnitude of  $\vb k$.  The first term (positive-frequency component) represents the emission of energy into the environment, while the second term (negative-frequency component) stands for the absorption of energy from the environment. It is evident that the summation over the momentum $\vb k$ yields distinct outcomes in various dimensions. After some algebraic manipulation, we obtain the following conclusions:
\( \al{&\text{1D}: \; S_{12}(\omega)=\cos(kd)S_{ii}(\omega),\\
       & \text{2D}: \; S_{12}(\omega)=J_0(kd)S_{ii}(\omega) ,\\
      & \text{3D}: \; S_{12}(\omega)= \frac{\sin (kd)}{kd} S_{ii}(\omega).  }\)                         
Here, $J_0(kd)$ represents the Bessel function of the first kind, which decays algebraically at large distances as $J_0(kd)\propto \cos(kd-\pi/4)/\sqrt{kd}$. From this, it is evident that the correlated noise is always bounded by the local noise, that is, $|S_{12}|\leq S_{ii}$.  Finally, we remark that we have assumed that the quasiparticle does not decay when traveling between the two spins.

\section{Time convolutionless master equations}\label{appendix2}
In Sec.~\ref{sec23} of the main text, we have presented the TCL master equations for the two-qubit system, which allow us to investigate the effects of correlated classical and quantum noise. In this section, we provide detailed derivations for the results presented in the main text. Firstly, in Appendix~\ref{appendixb1}, we recap the time-convolutionless projection operator method. Based on this method, we derive the master equation for pure-dephasing noise in Appendix~\ref{appendixb2}, and the master equation for pure-transverse noise in Appendix~\ref{appendixb3}.

\subsection{The time-convolutionless projection operator method}\label{appendixb1}
For the sake of self-consistency, here we briefly review the TCL master equation that we employ in the main text. Let us consider a system $S$ of our interest coupled to an environment $E$ that we will not keep track of. The dynamics of the combined system is governed by a microscopic Hamiltonian of the form: 
\( H=H_S+H_E+H_{\text{SE}},  \)
where $H_S$ and $H_E$ dictate the time evolution of the system $S$ and the environment $E$, respectively, and $H_{\text{SE}}$ describes the coupling between them. In the platform considered in the main text, $H_S$ is the two-qubit system that is of interest and $H_E$ is the environment that gives rise to local and correlated noise, leading to decoherence of qubits via the coupling $H_{\text{SE}}$. It is convenient to work in the interaction picture, where the density matrix of the combined system $\rho_{\text{tot}}(t)$ obeys the following equation of motion: 
\( \dv{\rho_{\text{tot}}(t)}{t} = \mathcal{L}(t)\rho_{\text{tot}}(t), \label{Eq:b2} \)
with the Liouville superoperator defined as $\mathcal{L}(t)\rho_{\text{tot}}(t)\equiv -i[H_{\text{SE}}(t), \rho_{\text{tot}}(t)]/\hbar$ and the coupling in interaction picture $H_{\text{SE}}(t)\equiv \text{exp}[i(H_S+H_E)t/\hbar] H_{\text{SE}}  \text{exp}[-i(H_S+H_E)t/\hbar]$.
We aim to derive the equation of motion for the reduced density matrix $\rho$ of the system $S$. To this end, we introduce a projection superoperator $\mathcal{P}$ that projects any density matrix $\rho_{\text{tot}}$ onto the system part of the Hilbert space: $\mathcal{P}\rho_{\text{tot}}\equiv \tr_E[\rho_{\text{tot}}]\otimes \rho_B$, where the trace is taken over the environment and $\rho_B$ is the initial state of the environment which we take to be the thermal state in the main text. We note that our goal is to obtain a closed equation for $\mathcal{P}\rho_{\text{tot}}$, which would give us the equation for $\rho$ naturally. 
Accordingly, a complementary superoperator $\mathcal{Q}$ can be defined by $\mathcal{Q}\equiv \mathcal{I}-\mathcal{P}$, with the identity operator $\mathcal{I}$, which projects on the irrelevant part of the density matrix. By applying the projection operators $\mathcal{P}$ and $\mathcal{Q}$ to the Liouville-von Neumann equation~\eqref{Eq:b2}, we arrive at
\(\al{ \dv{}{t}\mathcal{P}\rho_{\text{tot}}(t)&=\mathcal{P}\mathcal{L}(t) \mathcal{P} \rho_{\text{tot}}(t) +\mathcal{P} \mathcal{L}(t)\mathcal{Q}\rho_{\text{tot}}(t) ,\\
           \dv{}{t}\mathcal{Q}\rho_{\text{tot}}(t)&=\mathcal{Q}\mathcal{L}(t)\mathcal{P}\rho_{\text{tot}}(t)+\mathcal{Q}\mathcal{L}(t)\mathcal{Q}\rho_{\text{tot}}(t).  }\)
        The idea now is to solve for $\mathcal{Q}\rho_{\text{tot}}(t)$ formally and substitute it into the equation for $\mathcal{P}\rho_{\text{tot}}$, which results in a closed equation for $\mathcal{P}\rho_{\text{tot}}$. At this point, there are typically two ways to proceed, both of which give rise to exact but conceptually different equations of motion for the reduced density matrix. These methods are detailed in Ref.~\cite{Heinz}. One approach leads to the well-known Nakajima-Zwanzig equation, which is a time-non-local equation containing a memory kernel. In contrast, the second method yields a time-convolutionless master equation with the following form:
           \( \dv{}{t}\mathcal{P}\rho_{\text{tot}}=\mathcal{K}(t)\mathcal{P}\rho_{\text{tot}}(t),  \)
where $\mathcal{K}(t)$ is a time-local generator, known as the TCL generator, and we have assumed the initial state of the combined system takes the form of $\rho_{\text{tot}}(0)=\rho(0)\otimes \rho_B$ thus $\mathcal{Q}\rho_{\text{tot}}(0)=0$.   As customary, we assume $\tr_E[H_{\text{SE}}(t)\rho_E]=0$ (namely, the noise operator $E_i$ in the main text has vanishing mean in state $\rho_B$)         and also assume that  coupling between the system $S$ and the environment is sufficiently   weak. Thus we can truncate the TCL generator $\mathcal{K}(t)$ at second order, yielding: 
\( \mathcal{K}(t)=\int^t_0ds\; \mathcal{P}\mathcal{L}(t) \mathcal{L}(s)\mathcal{P}.  \)
Therefore, the TCL master equation for the reduced density matrix $\rho$ in the interaction picture takes the form of
\(  \dv{}{t}\rho(t)  =  \int^t_0ds\; \tr_E \big[\mathcal{L}(t) \mathcal{L}(s) \rho(t)\otimes \rho_B\big].  \label{B6}  \)
This equation serves as the starting point of our following derivations. 

\subsection{Master equation for pure-dephasing noise}\label{appendixb2}
In this subsection, we drive the TCL master equations for the two-qubit system in the absence of the coherent drive. In this scenario, the qubit-environment coupling is given by
\( H_{\text{SE}}=\sum_{i\in\{1,2\} } \sigma^z_i E_i,   \)
where $E_i$ are the noise operators acting on the Hilbert space of the environment. With this interaction, we can derive the master equation by utilizing the TCL equation~\eqref{B6}:
\(\al{ \dv{\rho(t)}{t}=-\frac{1}{\hbar^2} & \sum_{i,j}\int^t_0ds\;  \big[  S_{ij}(t-s) (\sigma
^z_i \sigma^z_j \rho - \sigma_j^z \rho \sigma_i^z) \\ 
     &  + S_{ij}(s-t) (\rho \sigma^z_i \sigma^z_j - \sigma^z_j \rho \sigma^z_i)    \big],   }\)
     where we have introduced the two-point correlation function of the noise operators $\tr_E[ E_i(t)E_j(s) \rho_B]=\tr_E[ E_i(t-s)E_j \rho_B]\equiv S_{ij}(t-s)$. Here we have used the fact that the initial state of the environment is the thermal state which is stationary. We note that terms with $i=j$ correspond to the standard local dephasing dynamics rooted in the local noise, whereas the terms with $i\neq j$ describe the collective  dynamics originated from the correlated noise. 
      We can cast the equation above into the following form: 
     \(\al{  \dv{\rho(t)}{t}=-\frac{1}{2\hbar^2} & \sum_{i,j}  \int^t_0ds\; \big[  S_{ij}(s)-S_{ij}(-s)  \big] [\sigma_i^z\sigma_j^z,\rho  ]   \\
                                     &+ \frac{1}{\hbar^2} \int^t_0 ds\;  \big[  S_{ij}(s)+S_{ij}(-s)  \big]  \mathcal{D}^z_{ij} [\rho] ,    } \)
                                     where the dissipator is defined as usual: $\mathcal{D}^z_{ij} [\rho] \equiv \sigma^z_j\rho \sigma_i^z -\{ \sigma_i^z\sigma_j^z,\rho \}/2$.
                                   We remark that there is also a term proportional to $\sigma_i^z$, which describes the induced Lamb shift and renormalizes the qubit energy spliting.  It is  rooted in the local noise  and we thus have neglected it since our main concern is the correlated noise.    Then we can conveniently write the master equation above into the following form: 
                                     \(   \dot{\rho}(t)=-i [ H_z(t), \rho(t)  ] +\sum_{i,j\in\{1,2\}} \mathcal{L}^z_{ij}(t)\rho(t).  \)
Below, we focus on the coherent part (which leads to unitary evolution) and the dissipative part (which gives rise to non-unitary dynamics), respectively.
                     
\textbf{Coherent interaction.}      The environment-induced coherent Ising interaction takes the form of $H_z(t)=\mathcal{J}^z(t)\sigma_1^z\sigma_2^z$ with the coupling being
                                  \( \al{  \mathcal{J}^z(t)  &=\frac{1}{i\hbar^2} \sum_{i\neq j}  \int^t_0 ds\;  \frac{S_{ij}(s)-S_{ij}(-s)}{2}  \\
                                                              &=\frac{1}{2\hbar^2}   \int^t_0ds\; \big[ \mathcal{G}^R_{12}(s)+ \mathcal{G}^R_{21}(s)     \big] ,   }  \label{eqb11}  \)
                                     where $\mathcal{G}_{12}^R(s)\equiv -i \Theta(s)\langle [E_1(s), E_2]\rangle$ represents the standard retarded Green's function, which indicates the retarded interaction between the two qubits mediated by the environment.  One can easily check that this coupling function is real-valued according to its definition above.    We aim to examine the effect of the correlated quantum and classical  noise. It is helpful to write the expression in a more suggestive form by utilizing $S_{ij}(t)=\int d\omega\,e^{-i\omega t} S_{ij}(\omega)/2\pi$, 
                                     \(  \mathcal{J}(t)=\frac{1}{\hbar^2} \int_{-\infty}^{\infty} \frac{d\omega}{2\pi} \big[ S_{12}(\omega) +S_{21}(\omega) \big] F_c(\omega, t),     \)
 with the filter function $F_c(\omega, t)= [\cos\omega t-1]/\omega$. We now further decompose the noise power spectral densities into classical and quantum components [see Eq.~\eqref{eq:cqnoise}], arriving at:
 \( \mathcal{J}(t)=\frac{2}{\hbar^2} \int^\infty_0 \frac{d\omega}{2\pi} \big[ S^Q_{12}(\omega) +S^Q_{21}(\omega)  \big] F_c(\omega, t), \label{eqb13}  \)
 which is the Eq.~\eqref{eq14} in the main text. It is clear that the correlated classical noise spectral densities cancel out, suggesting that this coherent Ising interaction is solely  determined by the correlated quantum noise. An alternative way to understand this is by referring to Eq.~\eqref{eqb11}. The coherent coupling arises solely from the commutators of the noise operators $E_i$. This implies that the coupling $\mathcal{J}^z$ vanishes when only classical noise is present, as $E_i$ can be treated as classical variables, and the commutators then vanish.

 \textbf{Dissipative evolution.}  The dissipative part is given by
 \( \mathcal{L}^z_{ij}(t) \rho=\gamma^z_{ij}(t) \mathcal{D}^z_{ij}\rho,\;\; \text{with}\;\;\gamma^z_{ij}(t)= \frac{1}{\hbar^2} \int^t_{-t} ds\; S_{ij}(s).  \)
It is clear that $\gamma^z_{ii}(t)$ is the local dephasing noise determined by the auto noise correlator $S_{ii}$, whereas $\gamma^z_{12}(t)$ stands for the correlated dephasing governed by the cross noise correlator $S_{12}$.
In terms of the noise spectral density, the pure-dephasing parameters can be written as
\(  \gamma^z_{ij}(t)= \frac{2}{\hbar^2}\int^\infty_{-\infty}\frac{d\omega}{2\pi}  S_{ij}(\omega) F_s(\omega, t),   \)
with the filter function $F_s(\omega, t)=\sin\omega t/\omega$. When we further decompose the noise spectral densities into quantum and classical components, we arrive at: 
\( \al{   \gamma^z_{ij}(t)=   \frac{2}{\hbar^2}  \int^\infty_0  \frac{d\omega}{2\pi} &  F_s(\omega, t) \big\{  [S^C_{ij} (\omega) +S^C_{ji}(\omega) ]   \\
                                   &  +      [S^Q_{ij} (\omega) -S^Q_{ji}(\omega) ]       \big\}   .  }  \label{eqb16}   \)
This is the Eq.~\eqref{eq15} in the main text by using the fact that $S_{ij}^*(\omega)=S_{ji}(\omega)$. First, we observe that the local dephasing rate $\gamma^z_{ii}$ is solely determined by the local classical noise $S^C_{ii}(\omega)$; the quantum component does not enter the local dephasing. In contrast, for the correlated dephasing process, the rate $\gamma^z_{12}$ is governed by both $\Re S^C_{12}(\omega)$ and $\Im S^Q_{12}(\omega)$. As we discussed in the main text, $\Im S^Q_{12}$ vanishes unless the spectrum of the quasiparticle in the environment $\omega_{\vb k}$ is asymmetric. Otherwise, in the case of symmetric environment, both local and correlated dephasing processes are dictated by the classical noise, and the quantum noise only leads to the coherent Ising interaction between qubits.

\subsection{Master equation for pure-transverse noise}\label{appendixb3}
In this subsection, we derive the TCL master equations for the two qubit system in the presence of coherent drives at resonance. In this scenario, the coupling between the system and the environment is given by
\( H_{\text{SE}}(t)=-\sum_i \hat{\sigma}_i^x E_i(t) =- \sum_i \big(  e^{i\Omega t} \hat{\sigma}_i^+ + e^{-i\Omega t} \hat{\sigma}_i^-  \big) E_i(t)   \)
in the interaction representation with $\hat{\sigma}^\pm\equiv (\hat{\sigma}^x\pm i\hat{\sigma}^y)/2$. With this interaction, we can derive the master equation:
\( \al{ \dv{\rho(t)}{t} =-\frac{1}{\hbar^2}\sum_{ij}\int^t_0ds\; \big[  e^{i\Omega s}  S_{ij}(s)(\hat{\sigma}^+_i \hat{\sigma}_j^- \rho - \hat{\sigma}^-_j \rho \hat{\sigma}^+_i)  \\ 
        + e^{-i\Omega s}S_{ij}(s) (\hat{\sigma}^-_i \hat{\sigma}_j^+ \rho - \hat{\sigma}^+_j \rho \hat{\sigma}^-_i  ) +e^{-i\Omega s} S_{ij}(-s) \\( \rho\hat{\sigma}^+_i \hat{\sigma}_j^-  - \hat{\sigma}^-_j \rho \hat{\sigma}^+_i )    +e^{i\Omega s} S_{ij}(-s) (\rho\hat{\sigma}^-_i \hat{\sigma}_j^+  - \hat{\sigma}^+_j \rho \hat{\sigma}^-_i)  \big]  .  } \)
We can then regroup the terms on the right hand side and rewrite the equation into the following compact form: 
\( \dot{\rho}(t)=-i [ H_{xy}(t)  ,\rho(t)] +\sum_{i,j\in\{1,2\}} \mathcal{L}_{ij}(t)\rho(t).  \label{B19} \)
This is the Eq.~\eqref{eq:mastereq} in the main text.   We  discuss below the time-dependent coherent interaction $H_{xy}(t)$ and the dissipative part, respectively. 

\textbf{Coherent interaction.} The coherent coupling induced by the environment is given by $H_{xy}(t)=\mathcal{J}(t) \hat{\sigma}_1^+\hat{\sigma}_2^- +\mathcal{J}^*(t) \hat{\sigma}_1^-\hat{\sigma}_2^+$ with time-dependent coupling strength being: 
\( \al{  \mathcal{J}(t) &=\frac{1}{2i\hbar^2} \int^t_0 ds\; \big[  e^{i\Omega s} S_{12}(s) -e^{-i\Omega s} S_{12}(-s) \\ & \;\;\;\;\;\;\;\;\;\;\;\;\;\;\;\;\;\;+e^{-i\Omega s} S_{21}(s)-e^{i\Omega s} S_{21}(-s)   \big] \\
&=\frac{1}{2\hbar^2} \int^t_0 ds\; \big[  \mathcal{G}^R_{12}(s) e^{i\Omega s} + \mathcal{G}^R_{21}(s) e^{-i\Omega s}   \big].    } \label{B20}   \)
We see that, again, the coupling can be expressed in terms of the retarded Green's function, suggesting that the coherent interaction is rooted in the correlated quantum noise, similar to the coherent Ising interaction we discussed before. This point becomes clear in the discussion below. Let us first express the coherent interaction in terms of the noise spectral density $S_{12}(\omega)$:
\(  \mathcal{J}(t)\!=\!\frac{1}{\hbar^2} \!\! \int^\infty_{-\infty} \!\!\frac{d\omega}{2\pi} \big[ S_{12}(\omega) F_c(\omega-\Omega,t) +S_{21}(\omega) F_c(\omega+\Omega,t)   \big] .   \)
We now further decompose the spectral density $S_{12}(\omega)$ into positive and negative part which can be expressed in terms of the quantum and classical noise.    We finally arrive at 
\( \mathcal{J}(t)\!\!=\!\!\frac{2}{\hbar^2}\!\!\! \int^\infty_0 \!\!\frac{d\omega}{2\pi} \big[ S^Q_{12}(\omega) F_c(\omega-\Omega, t) + S^Q_{21}(\omega) F_c(\omega+\Omega, t)  \big] , \)
which is Eq.~\eqref{eq18} in the main text. It is clear that this coherent coupling  takes a similar form as the Ising coupling $\mathcal{J}^z(t)$, and  at $\Omega\rightarrow 0$, we obtain $\mathcal{J}(t)=\mathcal{J}^z(t)$.
However, we should point out one crucial difference between them: the Ising coupling is always real whereas the coherent coupling $\mathcal{J}(t)$ is complex in general consisting of a real part that describes the symmetry exchange between the two qubits and a imaginary part that represents the Dzyaloshinskii–Moriya (antisymmetric exchange) interaction.

\textbf{Dissipative evolution.} The dissipator in the master equation~\eqref{B19} is given by the following Lindbladians: 
\(   \al{ \mathcal{L}_{ij}\rho &= \gamma_{ij}^\downarrow(t)\Big[ \hat{\sigma}_j^- \rho \hat{\sigma}_i^+ -\frac{1}{2}\{  \hat{\sigma}_i^+\hat{\sigma}_j^-,\rho    \}   \Big]   \\
   &\; +  \gamma_{ij}^\uparrow(t)\Big[ \hat{\sigma}_j^+ \rho \hat{\sigma}_i^- -\frac{1}{2}\{  \hat{\sigma}_i^-\hat{\sigma}_j^+,\rho    \}   \Big] ,}    \)
with rates being
\(\al{ &\gamma_{ij}^\downarrow(t) \!\!=\!\! \frac{1}{\hbar^2}\!\! \int^t_{-t}\!\!\! ds S_{ij}(\tau) e^{i\Omega \tau} \!\!=\!\! \frac{2}{\hbar^2} \!\!\int^{\infty}_{-\infty} \!\!\frac{d\omega}{2\pi} S_{ij}(\omega) F_s(\omega\!-\!\Omega, t), \\
  & \gamma_{ij}^\uparrow(t) \!\!=\!\! \frac{1}{\hbar^2}\!\! \int^t_{-t}\!\!\! ds S_{ij}(\tau) e^{-i\Omega \tau} \!\!=\!\! \frac{2}{\hbar^2} \!\!\int^{\infty}_{-\infty} \!\! \frac{d\omega}{2\pi} S_{ij}(\omega) F_s(\omega\!+\!\Omega, t).  }\label{B24}\)
  Here terms with $i=j$ stand for the local emission and absorption processes governed by the local noise spectral density $S_{ii}(\omega)$, whereas the terms with $i\neq j$ represent the correlated emission and absorption processes rooted in the cross noise spectral density $S_{12}(\omega)$. In the long time dynamics $\Omega t\gg 1$, the filter function $F_s(\omega\pm \Omega,t)$ approaches a delta function $\pi \delta(\omega\pm \Omega)$. In this case, the decay and absorption rates become time-independent and    we can approximate them with $\gamma^\downarrow_{ij}=S_{ij}(\Omega)/\hbar^2$ and $\gamma^\uparrow_{ij}=S_{ij}(-\Omega)/\hbar^2$. They are related by the Boltzmann factor, $\gamma^\downarrow_{ij} = e^{\beta \hbar \Omega}  \gamma^\uparrow_{ji}$. This is the detailed balance condition. Specifically, at low temperatures, the decaying process dominates and we can approximate $\gamma^\uparrow_{ij}\approx 0$. To further illustrate the effects of the classical and quantum noise, we express the  local and correlated decay rates in terms of the classical and quantum noise spectral densities, yielding: 
\( \al{  \gamma_{ij}^\downarrow(t)  \!\!&= \!\! \frac{2}{\hbar^2} \!\!\int^\infty_0\!\! \frac{d\omega}{2\pi}  \Big[ S_{ij}^C(\omega) F_s(\omega\!-\!\Omega, t) \!\!+\!\!S^C_{ji}(\omega) F_s(\omega\!+\!\Omega, t)  \Big]   \\
 &+\!\!  \frac{2}{\hbar^2} \!\!\int^\infty_0 \!\!\frac{d\omega}{2\pi} \Big[ S_{ij}^Q(\omega) F_s(\omega\!-\!\Omega,t) \!\!-\!\!S^Q_{ji}(\omega) F_s(\omega\!+\!\Omega,t) \Big].   }  \label{eqB25}  \)
 This is the equation~\eqref{eq19} in the main text. 
 The local and correlated absorption rate $ \gamma_{ij}^\uparrow(t) $ is given by the same expression above but with $\Omega\rightarrow -\Omega$. It is now clear that, in contrast to the pure dephasing dynamics, here the quantum noise leads to    both local and correlated decoherence. 
 We can similarly approximate the filter function with a delta function when $t\gg 1/\Omega$. We then arrive at $\gamma^\downarrow_{ij}=\big[  S^C_{ij}(\Omega) +S^Q_{ij}(\Omega)   \big]/\hbar^2$ and $\gamma^\uparrow_{ij}=\big[  S^C_{ji}(\Omega) -S^Q_{ji}(\Omega)   \big]/\hbar^2$, which is the equation~\eqref{eq:markovianrate} in the main text. From these expressions, we  conclude that the asymmetry between the decay and absorption processes is rooted in the quantum noise. In the absence of any quantum noise, decay and absorption occur with equal strength, corresponding to the infinite temperature limit.
 
 At this point, it is beneficial to recapitulate the dimensions of some critical parameters and functions that  are frequently used in the main text:
 \(\al{ [E_i]&=\text{energy}, \;\; [S_{ij}(t)]=\text{energy}^2, \;\; [F_{c,s}]=\text{time},  \\
          [\gamma_{ij}]&=\text{time}^{-1}, \;\;  [S_{ij}(\omega)]=\text{time}  \cdot \text{energy}^2, \\
          [\mathcal{J}^z]&=  [\mathcal{J}] =\text{time}^{-1}. }\)
and the $\sigma$ in the definition of the $1/f$ noise has the dimension of energy. 

\section{Pure dephasing dynamics} \label{appendixc}
In Sec.~\ref{sec_3} in the main text, we study the pure dephasing dynamics when the two qubits are subjected to correlated classical and quantum $1/f$ noise. Here, we solve the corresponding TCL master equation analytically and derive some results used in Sec.~\ref{sec_3}. For pure dephasing dynamics, it is convenient to work in the basis $\ket{a} \in \{\ket{\uparrow\uparrow}, \ket{\uparrow\downarrow}, \ket{\downarrow\uparrow}, \ket{\downarrow\downarrow}\}$ (eigenstates of $\sigma_1^z\otimes \sigma_2^z$). Let us denote the density matrix as $\rho(t)=\sum_{a,b} G_{ab}\ket{a}\bra{b}$. Then all density matrix elements are decoupled from each other with the following equation of motions: 
\(\al{   \dot{G}_{12} &=-2\big[  \gamma^z(t)-i \Im \gamma^z_{12}(t)    \big] G_{12} -2i\mathcal{J}^z(t) G_{12},     \\
           \dot{G}_{13} &=  -2\big[  \gamma^z(t)+ i \Im \gamma^z_{12}(t)    \big] G_{12}   - 2i\mathcal{J}^z(t) G_{13},  \\
           \dot{G}_{24} &=  -2\big[  \gamma^z(t)- i \Im \gamma^z_{12}(t)    \big] G_{24}  +2i\mathcal{J}^z(t) G_{24}  ,\\
            \dot{G}_{34} &=  -2\big[  \gamma^z(t)+ i \Im \gamma^z_{12}(t)    \big] G_{34}  +2i\mathcal{J}^z(t) G_{34}  ,          }   \)
which are dependent on the local classical noise and spatially correlated quantum noise. Here, we  emphasize that the local dephasing parameter $\gamma^z(t)$ is determined solely by local classical noise, whereas both the imaginary part of correlated dephasing $\Im \gamma^z_{12}$ and the Ising coupling $\mathcal{J}^z$ are rooted in the imaginary and real parts of the correlated quantum noise spectral density $S_{12}^Q$, respectively. The remaining two off-diagonal elements are solely determined by local and spatially correlated classical noise:
\(\al{ \dot{G}_{23}=-4\big[ \gamma^z(t)-\Re \gamma^z_{12}(t)   \big]G_{23}, \\ \dot{G}_{14}=-4\big[ \gamma^z(t)+\Re \gamma^z_{12}(t)   \big]G_{14}, }  \) 
where we recall that $\Re\gamma^z_{12}$ is dictated by the correlated classical noise only. Since all these equations are decoupled, one can easily solve them. Here we restrict to the scenario that we discuss in the main text: namely the correlated noise is comparable with the local noise $S^C_{12}(\omega)\approx e^{i\theta}S_{ii}^C(\omega)$ and we are in the quantum regime such that $S^Q\approx S^C$. Here, we assume the phase of the correlated noise spectral function is constant $\theta$ for simplicity.   We then have $\gamma^z_{12}=e^{i\theta}\gamma^z$. 

When the local classical noise is $1/f$, i.e. $S^C_{ii}(\omega)=2\pi \sigma^2/|\omega|$, we can evaluate the Ising coupling $\mathcal{J}^z$ defined by Eq.~\eqref{eqb13}:
\(   \mathcal{J}^z(t)=\frac{4}{\hbar^2}\int^\infty_0\frac{d\omega}{2\pi} \cos\theta \frac{2\pi \sigma^2}{\omega} F_c(\omega, t)  =- \frac{2\pi\sigma^2\cos\theta}{\hbar^2}t,   \)
and the pure-dephasing parameter $\gamma^z$ defined by Eq.~\eqref{eqb16}:
\(  \gamma^z(t)=\frac{4}{\hbar^2} \int^\infty_{\omega_l} \frac{d\omega}{2\pi} F_s(\omega, t) \frac{2\pi \sigma^2}{\omega} =\frac{4\sigma^2 t}{\hbar^2} [1-\text{Ci}(\omega_l t)].  \)
These two equations give us Eq.~\eqref{eq22} in the main text. 
Here, we have introduced  a low frequency cutoff $\omega_l$ which is set by the experimental measurement time and  the cosine integral function defined by: 
\(   \text{Ci}(x)=-\int^\infty_x dt \frac{\cos t}{t}= \gamma+\ln x+\sum_{k=1}^\infty \frac{(-x^2)^k}{2k (2k)!} ,    \)      
where $\gamma$ is Euler's constant. Since the dynamics that we are interested in occurs within a time much shorter compared to the measurement time, i.e. $\omega_l t\ll 1$, we can approximate $\text{Ci}(x)=\gamma+\ln x$ and we thus have $\gamma^z(t)\approx 4\sigma^2 t[1-\gamma - \ln (\omega_l t)]/\hbar^2$.  When we solve the master equation, it is convenient to  introduce two time-dependent functions,
\(\al{   \Gamma^z(t) & \equiv \int^t_0ds\; \gamma^z(s)\approx  \frac{\sigma^2t^2}{\hbar^2}\big[ 3-2\gamma -2\ln (\omega_l t) \big],  \\
   V(t) & \equiv  \int^t_0 ds \mathcal{J}^z(s)=- \frac{\pi \sigma^2 t^2\cos\theta}{\hbar^2}.    }\)
   All elements can be expressed in terms of these two functions and are given by the following expressions:
\(\al{  G_{12}(t) & = G_{12}(0)\exp[-2(1-i\sin\theta) \Gamma^z(t)-2iV(t)  ],\\
           G_{13}(t) & = G_{13}(0)\exp[-2(1+i\sin\theta) \Gamma^z(t)-2iV(t)  ],\\ 
           G_{24}(t) & = G_{24}(0)\exp[-2(1-i\sin\theta) \Gamma^z(t)+2iV(t)  ],\\
           G_{34}(t) & = G_{34}(0)\exp[-2(1+i\sin\theta) \Gamma^z(t)+2iV(t)  ],\\ 
           G_{23}(t) & = G_{23}(0)\exp[-4(1-\cos\theta) \Gamma^z(t)  ],\\
             G_{14}(t) & = G_{14}(0)\exp[-4(1+\cos\theta) \Gamma^z(t)  ].   }\)
             With the explicit expression of the reduced density matrix $\rho(t)$, one can evaluate the entanglement of the two-qubit system as a function of time with arbitrary initial state, for example the two plots we show in Fig.~\ref{fig3} in the main text.

\section{Markovian limit}\label{appendixd}
In Sec.~\ref{seciv} of the main text, we discuss the dynamics of two qubits subjected to coherent drives with Markovian noise. In this section, we provide detailed derivations of some results used in the main text. In Sec~\ref{appendixd1}, we introduce the concurrence as a measure of the entanglement of the two qubits in the symmetrized and antisymmetrized basis. We then present an analytical solution for the master equation in the absence of the DM interaction in Sec.~\ref{appendixd2}. In Sec.~\ref{appendixd3}, we discuss the two qubit dynamics in the absence of the symmetric exchange interaction. In Sec.~\ref{appendixd4}, we present a study of the two qubit system when both the symmetric exchange and DM interactions are present. Finally, in Sec.~\ref{appendixd5}, we investigate the long-term dynamics of the two qubit system at finite temperatures.

The dynamics of the two-qubit system is governed by the master equation~\eqref{B19}. In the case of Markovian noise, one can extend the time to infinity in the expression of the coherent coupling $\mathcal{J}$ given by Eq.~\eqref{B20} and decay and absorption rates $\gamma^\downarrow_{ij}(t), \gamma^\uparrow_{ij}(t)$ given by Eq.~\eqref{B24}. We then have the following expressions: 
\( \mathcal{J}=\frac{ \mathcal{G}^R_{12}(\Omega) +  \mathcal{G}^R_{21}(-\Omega)  }{2\hbar^2}, \gamma^\downarrow_{ij}=\frac{S_{ij}(\Omega)}{\hbar^2},  \gamma^\downarrow_{ij}=\frac{S_{ij}(-\Omega)}{\hbar^2}.   \)
One important relation is $\gamma^\downarrow_{ij}=e^{\beta \hbar \Omega}\gamma^\uparrow_{ji}$, which allows us to gauge out the phase of $\gamma^\downarrow_{ij}$ (and also $\gamma^\uparrow_{ij}$) and absorb the phase into the coherent coupling $\mathcal{J}$. Therefore, we will assume  that $\gamma^\downarrow_{ij}$ and $\gamma^\uparrow_{ij}$ are positive-valued while $\mathcal{J}$ is complex in general. 

It is convenient to work in the symmetrized and antisymmetrized basis: $\ket{a}\in\{ \ket{\uparrow\uparrow}, \ket{T}, \ket{S}, \ket{\downarrow\downarrow}  \}$, where $\ket{T}$ and $\ket{S}$ are the standard triplet and singlet states. In this basis, let us denote the density matrix as 
\( \al{  \rho=& G_t\ket{T}\bra{T} +G_s \ket{S} \bra{S} +G_{11} \ket{\uparrow\uparrow}\bra{\uparrow\uparrow} +G_{44} \ket{\downarrow\downarrow}\bra{\downarrow\downarrow} \\
                  &  + G_{ts}\ket{T}\bra{S} +G_{st}\ket{S}\bra{T} +\Delta \rho,    } \)
       where $\Delta\rho$ stands for other off-diagonal elements. As we show below the dynamics of the density matrix elements that we introduce in the above expression are closed, we will  explore the dynamics  within this subspace.         
 From the master equation~\eqref{B19}, we deduce the equations of motion for   the diagonal elements:
               \( \al{ &\dot{G}_{11}=-2\gamma^\downarrow G_{11}  + ( \gamma^\uparrow + \gamma^\uparrow_{12}  ) G_t  + ( \gamma^\uparrow - \gamma^\uparrow_{12}  ) G_s  ,  \\
                   & \dot{G}_{44}=( \gamma^\downarrow + \gamma^\downarrow_{12}  )G_t + (\gamma^\downarrow -\gamma^\downarrow_{12} )G_s -2\gamma^\uparrow G_{44},  \\
                         &\dot{G}_t=-2\mathcal{D} x +(\gamma^\downarrow +\gamma^\downarrow_{12}) G_{11} - (\gamma^\downarrow+ \gamma^\downarrow_{12} )G_{t}   \\
                           &\;\;\;\;\; \;\;\;\;\; \;\;\;\;\;\;\;\;\;\; \;\;\;\;\;  + ( \gamma^\uparrow + \gamma^\uparrow_{12}  )G_{44} -( \gamma^\uparrow + \gamma^\uparrow_{12}  ) G_t   , \\
                          & \dot{G}_s=2\mathcal{D}x   +(\gamma^\downarrow -\gamma^\downarrow_{12}) G_{11} - (\gamma^\downarrow- \gamma^\downarrow_{12} )G_{s}  \\
                          & \;\;\;\;\; \;\;\;\;\; \;\;\;\;\; \;\;\;\;\; \;\;\;\;\;  +( \gamma^\uparrow - \gamma^\uparrow_{12}  )G_{44}  -( \gamma^\uparrow - \gamma^\uparrow_{12}  )G_s.   } \label{D3}  \)   
Here are some remarks regarding these equations. Firstly, we denote the local decay (absorption) rates as $\gamma^{\downarrow,\uparrow}\equiv \gamma^{\downarrow,\uparrow}_{ii}$, which must be larger compared to their nonlocal counterparts as guaranteed by the complete positivity of the dynamics. Secondly, we note that the change rate of the summation of the diagonal elements vanishes, as expected, suggesting that the evolution is trace-preserving. Thirdly, since all the basis states are eigenstates of the term $\propto \mathcal{J}_s$ of the coherent Hamiltonian, this term does not contribute to the dynamics of the diagonal elements and only results in the oscillation of off-diagonal elements $G_{ts}$. Lastly, since the term proportional to $\mathcal{D}$ breaks the parity, it relates the dynamics $G_t$ to $G_s$.

     Let us introduce $G_{ts}\equiv x+iy$ for convenience.   We can also write down the equation for the off-diagonal term: 
                          \( \al{ & \dot{x}=-(\gamma^\downarrow  +\gamma^\uparrow )x+2\mathcal{J}_s y +\mathcal{D}(G_t-G_s), \\
                          & \dot{y}=-(\gamma^\downarrow +\gamma^\uparrow ) y -2\mathcal{J}_s x.    }\)
      As anticipated,  these equations are closed.  Our discussion below will focus on this closed subspace.

   \subsection{Concurrence in symmetrized and antisymmetrized basis}   \label{appendixd1}
        For a pure bipartite state $\rho_{\text{AB}}= \ket{\psi_{\text{AB}}}\bra{\psi_{\text{AB}}}$, we usually adopt the von Neumann entropy as the entanglement measure: $S(\ket{\psi_{\text{AB}}})\equiv -\tr\rho_A\ln \rho_A=-\tr\rho_B\ln \rho_B$.  For a general mixed state $\rho_{\text{AB}}$, this von-Neumann entropy is no longer a good measure since the classical mixture in $\rho_{\text{AB}}$ will have a nonzero contribution. We will adopt entanglement of formation as our entanglement measure.

The entanglement of formation is defined as
\( E_F(\rho_{\text{AB}}) \equiv \text{min} \sum_i p_i\, S(\ket{\psi^i_{\text{AB}}}),   \)
where the minimum is taken over all possible decompositions of $\rho_{\text{AB}}=\sum_i p_i \ket{\psi^i_{\text{AB}}}\bra{\psi^i_{\text{AB}}}$ and $S(\ket{\psi^i_{\text{AB}}})$ is the von Neumann entropy of the pure state $\ket{\psi^i_{\text{AB}}}$. Physically, $E_F(\rho_{\text{AB}})$ is the minimum amount of pure state entanglement needed to create the mixed state. This is extremely difficult to evaluate in general since we need to try all the decompositions. Quite remarkably an explicit expression of $E_F(\rho_{\text{AB}})$ is given when both $A$ and $B$ are two-state systems (qubits). This exact formula is based on the often used two-qubit concurrence, which is defined as~\cite{PhysRevLett.80.2245} 
\( \mathcal{C}(\rho)=\text{max} \{ 0,\lambda_1-\lambda_2-\lambda_3-\lambda_4 \}, \label{concurrence} \)
where $\lambda_i$'s are, in decreasing order, the square roots of the eigenvalues of the matrix $\rho (\sigma_y\otimes \sigma_y)\rho^*(\sigma_y\otimes \sigma_y)$, where $\rho^*$ is the complex conjugate of $\rho$. The entanglement of formation is then given by~\cite{PhysRevLett.80.2245}
\(   E_F(\rho)= h\Big( \frac{1+\sqrt{1-\mathcal{C}^2}}{2}  \Big),  \)
with $h(x)=-x\log_2x-(1-x)\log_2(1-x)$.
$E_F(\rho)$ is monotonically increasing and ranges from 0 to 1 as $\mathcal{C}(\rho)$ goes from 0 to 1, so that one can take the concurrence as a measure of entanglement in its own right.  When we write the density matrix as: 
\( \rho=\sum_{i,j\in\{\uparrow,\downarrow\}}G_{ij}\ket{ij}\bra{ij} + G_{23}\ket{\uparrow\downarrow}\bra{\downarrow\uparrow}+ G_{32}\ket{\downarrow\uparrow}\bra{\uparrow\downarrow}, \)
                  and other off-diagonal elements vanish. The concurrence can be shown to be
                  \(  \mathcal{C}[\rho]=2 \max\big\{0, |G_{23}| -\sqrt{G_{11}G_{44}} \big\}.   \)
                  On the other hand, we know the subspace spanned by $\{\ket{\uparrow\downarrow}, \ket{\downarrow\uparrow}\}$ is linked to the subspace spanned by $\{ \ket{T}, \ket{S} \}$ through the following relation:
                  \(  \hat{\rho}= U^\dagger \rho U,\;\; \text{with}\; \;U=\frac{1}{\sqrt{2}} \mqty[1 & 1\\ 1& -1],    \)
                  where we note that $U$ is both hermtian and unitary. Then we can easily show that: 
                  \( G_{23}= \frac{G_t-G_s}{2} -i\Im G_{ts}.  \)
                  Therefore, in the basis that we used in the main text, the concurrence is given by: 
                  \(  \mathcal{C}[\rho]=\max \big\{ 0, | G_t-G_s -2i \Im G_{ts}  | -2\sqrt{G_{11}G_{44}}  \big\} .   \label{D12} \)
At zero temperature, the probability of staying in the state $\ket{\uparrow\uparrow}$ vanishes $G_{11}=0$ if the initial state has zero probability in the $\ket{\uparrow\uparrow}$ state. Consequently, the concurrence is further reduced to the following simple expression: 
           \(  \mathcal{C}[\rho]= | G_s-G_t +2i \Im G_{ts}  |,  \)
                  which gives us Eq.~\eqref{eq28} in the main text.

              Before we close this subsection, we highlight a quick and convenient method to evaluate the lower bound of entanglement of formation for a generic mixed state $\rho$~\cite{burkard2003lower,bennett1996mixed}. It is known that, for the Werner states~\cite{werner1989quantum}
                 \(   \rho_F=F\ket{S}\bra{S} + \frac{1-F}{3} \Big(  \ket{T}\bra{T} + \sum_{i=\pm} \ket{\Phi_i}\bra{\Phi_i} \Big),   \)
                  the entanglement of formation is determined by the singlet fidelity $F=\bra{S}\rho_F\ket{S}$. Here $\ket{S}, \ket{T}$ are singlet and triplet states, and $\ket{\Phi_\pm}\equiv( \ket{\uparrow\uparrow} \pm \ket{\downarrow\downarrow}  )/\sqrt{2}$ are the other two Bell states.  The entanglement of formation of $\rho_F$ is given by the following function~\cite{burkard2003lower,bennett1996mixed}:
                  \(  H(F)=\begin{cases}    h[1/2+\sqrt{F(1-F)}], &  \text{for}\; 1/2<F\leq 1;\\
                        0, &  \text{for}\; 0\leq F\leq 1/2.
                  \end{cases}   \) 
                 Here,  $h(x)\equiv -x\log_2x-(1-x)\log_2(1-x)$. While  the entanglement of formation of a generic mixed state $\rho$ is not completely determined by the singlet fidelity $F$ (we need to evaluate the relatively complicated quantity concurrence as we discussed above), the entangelment of formation of the corresponding Werner state [with $F=\bra{S}\rho\ket{S}$] provides a lower bound on $E_F(\rho)$~\cite{burkard2003lower,bennett1996mixed}: 
                 \(  H(F)\leq E_F(\rho).  \)
            In fact, this lower bound can be improved by  setting $F=\max \bra{e}\rho\ket{e}$, where the maximum is over all completely entangled states $\ket{e}$~\cite{bennett1996mixed,burkard2003lower}. One can interpret $F$ as the ``fully entangled fraction" of state $\rho$. Therefore, if $F>1/2$ for a given state $\rho$, it is entangled. However, it is important to note that obtaining $F\leq 1/2$ does not conclusively prove the absence of entanglement, as $H(F)$ only provides a lower bound. One example is 
            \( \rho_0=\frac{1}{2}\ket{\uparrow\uparrow}\bra{\uparrow\uparrow} + \frac{1}{2}\ket{S}\bra{S}.  \)
         We find $F=\bra{S}\rho_0\ket{S}=1/2$, but it is important to note that this state is entangled [with nonzero concurrence $\mathcal{C}(\rho_0)=1/2$], as it cannot be constructed from unentangled pure states.

                  \begin{figure}
	\centering\includegraphics[width=0.98\linewidth]{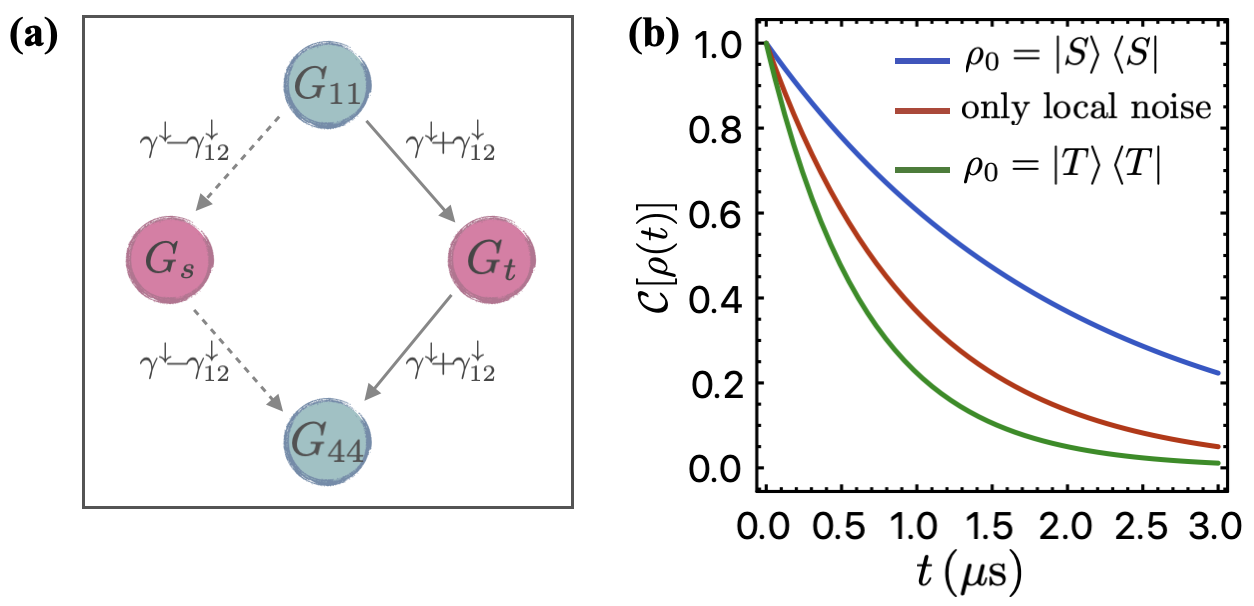}
	\caption{(a) The decay of the state $\ket{\uparrow\uparrow}$ to $\ket{\downarrow\downarrow}$ can occur through two pathways. The first pathway is via the triplet state $\ket{T}$, with a decay rate of $\gamma^\downarrow+\gamma^\downarrow_{12}$. The second pathway is via the singlet state $\ket{S}$, with a decay rate of $\gamma^\downarrow-\gamma^\downarrow_{12}$.  (b) When the DM interaction $\mathcal{D}$ is absent (i.e., $\mathcal{D}=0$), the entanglement between two qubits decays at a faster rate (the green curve) if the initial state is a triplet state $\ket{T}$, or decays at a lower rate (the blue curve) if the initial state is a singlet state $\ket{S}$, compared to the case when there is only local noise (the red curve, two initial states decay at the same rate). We assume a local decay rate of $\gamma^\downarrow=1\text{ GHz}$, and a correlated decay rate of $\gamma^\downarrow_{12}=0.5\text{ GHz}$.  }
	\label{figsmf1}
\end{figure}
                  
           \subsection{Symmetric exchange interaction}         \label{appendixd2}
         In this subsection and the next subsection, we restrict our discussion to the quantum regime where the absorption rates vanish $\gamma^\uparrow=\gamma^\uparrow_{12}=0$. In this scenario, the equations of motion for the density matrix elements are reduced to: 
              \( \al{ &\dot{G}_{11}=-2\gamma^\downarrow G_{11}, \,\; \dot{G}_{44}=( \gamma^\downarrow + \gamma^\downarrow_{12}  )G_t + (\gamma^\downarrow -\gamma^\downarrow_{12} )G_s,  \\
                         &\dot{G}_t=-2\mathcal{D} x +(\gamma^\downarrow +\gamma^\downarrow_{12}) G_{11} - (\gamma^\downarrow+ \gamma^\downarrow_{12} )G_{t}, \\
                          & \dot{G}_s=2\mathcal{D}x   +(\gamma^\downarrow -\gamma^\downarrow_{12}) G_{11} - (\gamma^\downarrow- \gamma^\downarrow_{12} )G_{s} \\
                           & \dot{x}=-\gamma^\downarrow  x+2\mathcal{J}_s y +\mathcal{D}(G_t-G_s),\;\; \dot{y}=-\gamma^\downarrow y -2\mathcal{J}_s x.   }  \label{eqd14}  \)   
                 Notably,  the state $\ket{\uparrow\uparrow}$ can decay to $\ket{\downarrow\downarrow}$ in two ways. The first way is through the triplet state, where we have the so-called superradiance with the decay rate $\gamma^\downarrow +\gamma^\downarrow_{12} $. The second way is through the singlet state, where we have the so-called subradiance with the decay rate $\gamma^\downarrow -\gamma^\downarrow_{12}$. This is sketched in Fig.~\ref{figsmf1} (a). This also  suggests that the singlet state and the triplet state decay at different rates in the presence of  correlated noise, which is illustrated in Fig.~\ref{figsmf1} (b). 

 When we turn off the DM interaction, $\mathcal{D}=0$, and assume the initial condition $\rho_0=\ket{\uparrow\downarrow}\bra{\uparrow\downarrow}$ (namely, $G_{t}=G_s=G_{ts}=1/2$), we first easily see that $G_{11}=0$ for all times.  Then the equations for $G_t$ and $G_s$ are decoupled from other elements (due to the absence of the parity-breaking interaction $\mathcal{D}$): 
           \(  \dot{G}_{t} =-( \gamma^\downarrow + \gamma^\downarrow_{12} )G_t, \;\;  \dot{G}_{s} =-( \gamma^\downarrow - \gamma^\downarrow_{12} )G_s,    \)
           which yield $G_t(t)=\text{exp}  [-(  \gamma^\downarrow + \gamma^\downarrow_{12} )t ]/2$ and $G_s(t)=\text{exp} [-(  \gamma^\downarrow - \gamma^\downarrow_{12} )t ]/2.$ By using the normalization $\tr \rho=1$, we conclude that: \(G_{44}(t) = 1-e^{-\gamma^\downarrow t} \cosh(\gamma^\downarrow_{12}t).  \)     
           For the off-diagonal element $G_{ts}$, we note that its real and imaginary parts $x$ and $y$ are coupled to each other.    To this end, it is helpful to introduce $X(t)=x(t)e^{\gamma^\downarrow t}$ and   $Y(t)=y(t)e^{\gamma^\downarrow t}$. The coupled equations then can be recast into the following compact form: 
           \( i \dv{}{t} \vb* \psi= M\vb* \psi,   \)
           with $\vb*\psi\equiv ( X, Y  )^T$ and $M=-2\mathcal{J}_s \sigma_y$, which is solved by $\vb*\psi(t)=\text{exp}(-iMt)\vb*\psi(0)$. By utilizing $\text{exp}(-iMt)=\cos(2\mathcal{J}_s t) +\sigma_y \sin(2\mathcal{J}_s t)$, we arrive at $X(t)=\cos(2\mathcal{J}_s t)/2$ and $Y(t)=-\sin(2\mathcal{J}_s t)/2$, where we have used the initial condition $X(0)=1/2$ and $Y(0)=0$. Therefore, we have \(G_{ts}(t)= \frac{\text{exp}[  -( \gamma^\downarrow +2i\mathcal{J}_s )t  ]}{2} .\)
           Thus, we have solved the master equation for the two-qubit system. 
    In this case, we can write down the entanglement (concurrence) of the two qubits as a function of time: 
    \( \al{\mathcal{C}[\rho(t)]& = | G_s-G_t +2i \Im G_{ts}  | \\
                                           &=e^{-\gamma^\downarrow t} \sqrt{ \sinh^2\gamma_{12}^\downarrow t +\sin^2(2\mathcal{J}_s t)  }.   }\)
This is Eq.~\eqref{eq30} in the main text. 

     \begin{figure}
	\centering\includegraphics[width=0.84\linewidth]{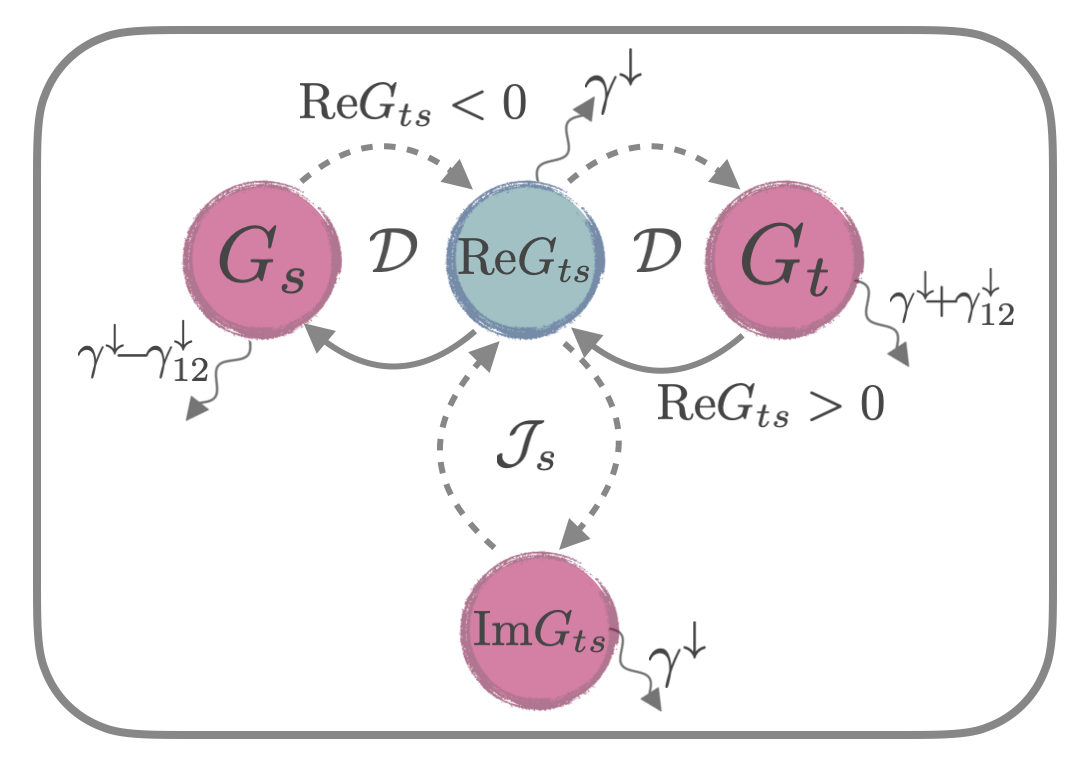}
	\caption{ Coupled dynamics of the populations in singlet state $G_s$ and triplet state $G_t$ governed by Eq.~\eqref{eqd14}. The DM interaction, which breaks the parity symmetry and is assumed to be positive-valued $\mathcal{D}>0$ for concreteness, couples the two states to each other. The flow of population from the triplet state to singlet state occurs when $\Re G_{ts}$ is positive and in the opposite direction when $\Re G_{ts}$ is negative. The dynamics of $\Re G_{ts}$ is determined by the relative populations of the singlet and triplet states, and is also coupled to $\Im G_{ts}$ via the symmetric exchange coupling $\mathcal{J}_s$. }
	\label{figsmf2}
\end{figure} 

 \begin{figure*}
	\centering\includegraphics[width=0.98\linewidth]{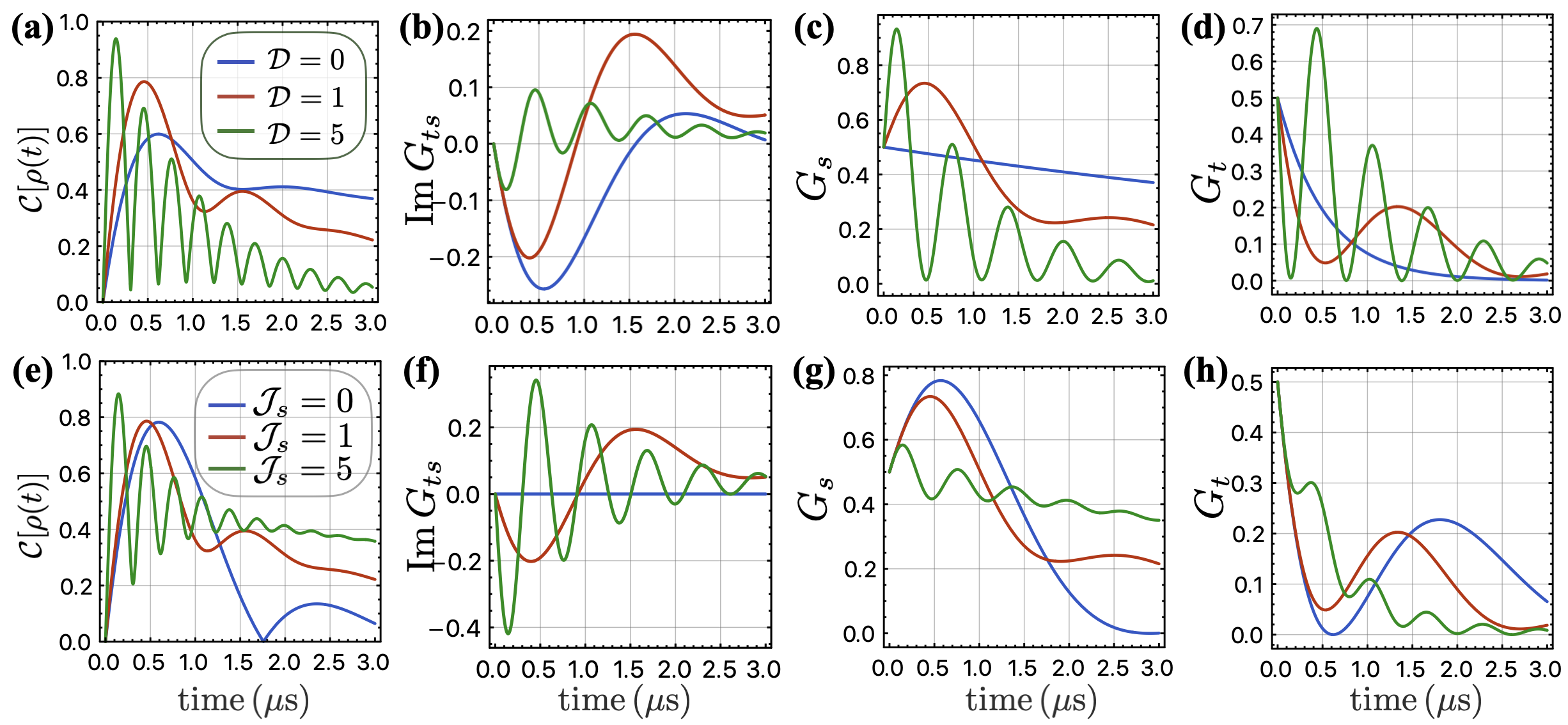}
	\caption{Entanglement dynamics of resonantly driven qubits with initial state $\ket{\uparrow\downarrow}$ in the presence of both symmetric exchange and DM interaction. (a)-(d) Qubits  dynamics with the fixed symmetric exchange coupling $\mathcal{J}_s=\gamma^\downarrow$ and varying DM interaction $\mathcal{D}$.   (a) The entanglement measured by the concurrence $\mathcal{C}[\rho(t)] $ between two qubits is plotted as a function of time for varying strengths of $\mathcal{D}$. As $\mathcal{D}$ increases, the maximum entanglement also increases and the oscillation frequency of entanglement becomes faster.
(b) The imaginary part of $G_{ts}$ is shown as a function of time, which exhibits faster oscillations as $\mathcal{D}$ increases. Furthermore, the magnitude of $\Im G_{ts}$ is suppressed with an increase in $\mathcal{D}$.
(c) The time-dependent function $G_s$ is plotted, which displays a larger value at short times for higher values of $\mathcal{D}$, but has a smaller value overall with increasing $\mathcal{D}$.
(d) The time-dependent function $G_t$ is plotted, which exhibits both faster oscillations and larger magnitudes with increasing $\mathcal{D}$.   (e)-(h) Qubits  dynamics with the fixed DM interaction $\mathcal{D}=\gamma^\downarrow$, and varying symmetric exchange coupling $\mathcal{J}_s$.(e) The entanglement between two qubits is plotted as a function of time, and it exhibits faster oscillations for larger values of $\mathcal{J}_s$. Moreover, as $\mathcal{J}_s$ increases, the magnitude of entanglement increases, but the amount of increase is smaller than in (a).
(f) The imaginary part of $G_{ts}$ is shown as a function of time, which exhibits faster oscillations and larger amplitude as $\mathcal{J}_s$ is increased.
(g) The time-dependent function $G_s$ is plotted and  displays faster oscillations for larger $\mathcal{J}_s$. Additionally, the magnitude of $G_s$ is suppressed at shorter times but achieves a larger value at longer times as $\mathcal{J}_s$ is increased.
(h) The time-dependent function $G_t$ is plotted, which exhibits faster oscillations and a suppressed magnitude as $\mathcal{J}_s$ is increased.    Parameters used in all figures: $\gamma^\downarrow=1\,\mu \text{s}^{-1}$ and $\gamma^\downarrow_{12}=0.9 \gamma^\downarrow$.  }
	\label{figsmf3}
\end{figure*}

           \subsection{Dzyaloshinskii–Moriya interaction} \label{appendixd3}
            Here, we turn off the symmetric exchange interaction $\mathcal{J}_s=0$, and assume the initial state to be $\ket{\uparrow\downarrow}$. Similarly to the case above, we conclude that $G_{11}(t)=0$. It is also clear that, since  $\dot{y}=-\gamma^\downarrow y$, we have $y(t)=0$ with the initial condition we assumed. Then all the equations are reduced to 
           \(  \dv{}{t}\bar{\vb* \psi} =\bar{M}\bar{\vb* \psi},    \)
           with $\bar{\vb* \psi} =(G_t, G_s, x)^T$ and 
           \(  M=\mqty[  -\Gamma_S  & 0 & -2 \mathcal{D}  \\ 0 & -\Gamma_A  & 2\mathcal{D}  \\ \mathcal{D}  & -\mathcal{D}  & -\gamma^\downarrow  ],   \)
           where $\Gamma_{S,A}=\gamma^\downarrow \pm \gamma^\downarrow_{12} $. To solve these three coupled differential equations, we  convert them into a single third order differential equation. To this end, we first introduce $\bar{G}_t=G_te^{\gamma^\downarrow t}, \bar{G}_s=G_se^{\gamma^\downarrow t}, X= xe^{\gamma^\downarrow t}$. Then these equations are reduced to the following form: 
           \( \al{  & \dot{\bar{G}}_t = -\gamma^\downarrow_{12} \bar{G}_t  -2\mathcal{D}X, \\
                      & \dot{\bar{G}}_s =  \gamma^\downarrow_{12} \bar{G}_s  +2\mathcal{D}X, \\
                       & \dot{X}=\mathcal{D} ( \bar{G}_t -\bar{G}_s  ).  }\)
                       By taking one more time derivative for $X$, we obtain a second order differential equation for $X$ but not closed: 
                       \(   \ddot{X}=-\mathcal{D}\gamma^\downarrow_{12} ( \bar{G}_{t} +\bar{G}_s ) -4\mathcal{D}^2X.      \)
           We note that, when  taking one more derivative, we have a closed third order differential equation for $X$: 
           \(   \dddot{X} +(4\mathcal{D}^2 -\gamma^{\downarrow 2}_{12}  ) \dot{X}=0,  \)
           which can be solved easily together with the initial conditions: $\dot{X}(0)=0, \ddot{X}(0)=-\mathcal{D}\gamma^\downarrow_{12}-2\mathcal{D}^2.$ For example, when $4\mathcal{D}^2 >\gamma^{\downarrow 2}_{12} $, we have: 
           \(  \dot{X}(t)=- \frac{\mathcal{D} \gamma^\downarrow_{12} +2\mathcal{D}^2  }{\omega_r} \sin\omega_r t,      \)
           with $\omega_r= \sqrt{4\mathcal{D}^2 -\gamma^{\downarrow 2}_{12}  }$. It is then straightforward to obtain the analytic expressions for all the density matrix elements. Focusing on the dynamics of  the entanglement between the qubits, one can easily write down its expression according to $ \mathcal{C}[\rho(t)]= |G_s-G_t|$ once we have the explicit expression for the density matrix. Distinct two-qubit entanglement dynamics can be achieved with different values of the DM interaction $\mathcal{D}$. We also remark that the interplay between the DM interaction and the dissipative process can also lead to intriguing physics in classical dynamics~\cite{zou2023dissipative,yu2023non}.

           On the other hand, it is helpful to write down the equation of motion for the entanglement directly and study how the quantum coherence in the system evolves. To this end, we introduce $\mathcal{C}_R\equiv G_s-G_t$ and we  derive a differential equation for it. 
           We note that it is linked to $\dot{X}$ via: $\mathcal{C}_R=- \dot{X} e^{-\gamma^\downarrow t}/ \mathcal{D}$, from which we conclude that: 
           \(   \ddot{ \mathcal{C} }_R+2\gamma^\downarrow \dot{\mathcal{C}}_R +(\gamma^{\downarrow 2} -\gamma^{\downarrow 2}_{12} +4\mathcal{D}^2 ) \mathcal{C}_R=0.   \)
           This is the equation~\eqref{eq31} in the main text, from which one can solve for the entanglement dynamics directly. 
          
           \subsection{Symmetric and DM interactions}\label{appendixd4}

        \begin{figure}
	\centering\includegraphics[width=0.7\linewidth]{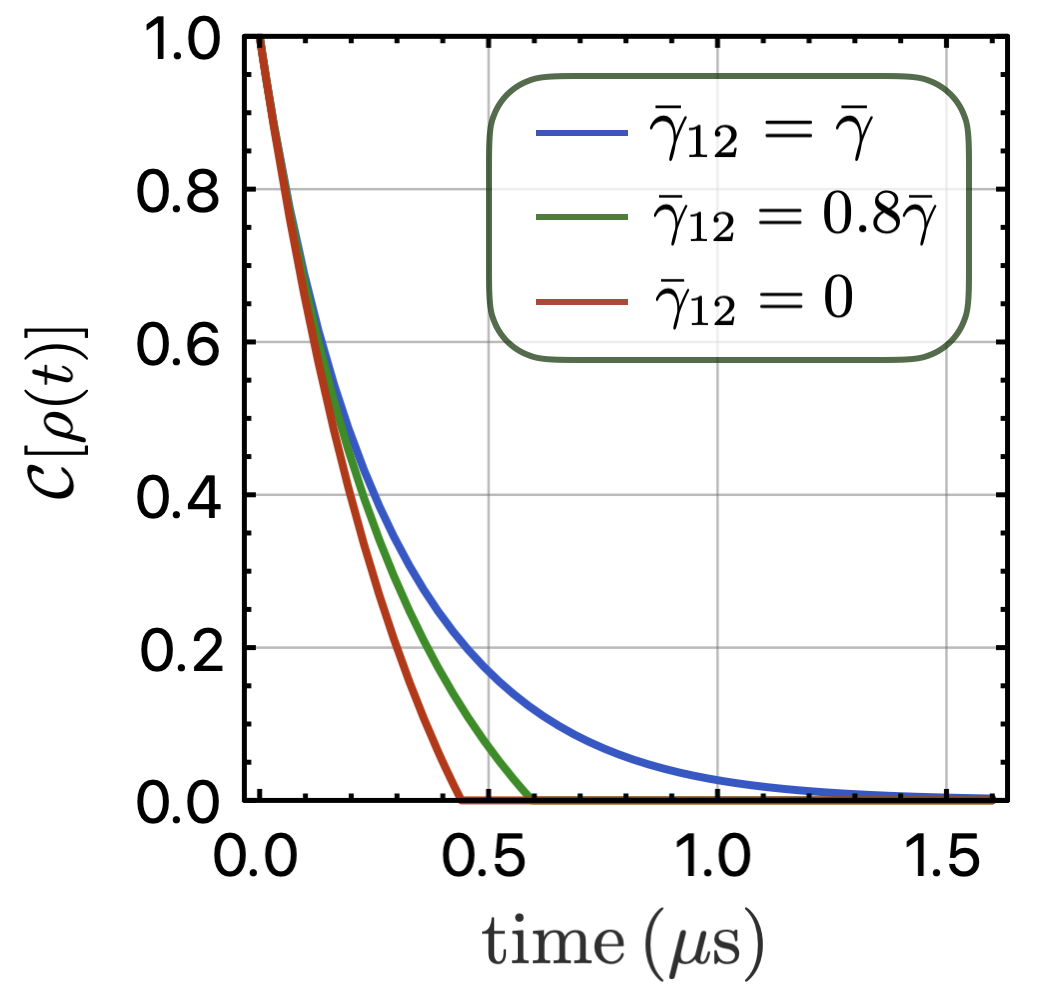}
	\caption{Dynamics of entanglement between two qubits as a function of time in the presence of pure classical Markovian noise. The system is initialized in the Bell state $\ket{\psi_0}=(\ket{\uparrow\downarrow}+i\ket{\downarrow\uparrow})/\sqrt{2}$. The curves correspond to the entanglement decay under different spatially correlated noise conditions. It is observed that the presence of correlated classical noise results in a slight modification of the decoherence rate, but does not introduce any new features. Parameters used in the plot:  $\bar{\gamma}=1\,\mu \text{s}^{-1}$. }
	\label{figsmf4}
\end{figure} 

We illustrate the effects of the symmetric exchange and the DM interaction, respectively, by solving the master equation for the two qubits analytically. This allows us to extrapolate the scenario when both are present. In this case, all four important elements are coupled to each other, as shown in Fig.~\ref{figsmf2}. The populations of the two states, $G_t$ and $G_s$, are coupled to each other by the DM interaction (assumed to be positive for concreteness), which breaks the parity symmetry. Population flows from the triplet state to the singlet state when $\Re G_{ts}$ is positive, and in the opposite direction when it is negative. The dynamics of $\Re G_{ts}$ depend on the relative populations of the singlet and triplet states and are further coupled to $\Im G_{ts}$ through the symmetric exchange coupling $\mathcal{J}_s$.

In Fig.~\ref{figsmf3}, we present the entanglement dynamics in two different scenarios. In the first case, we fix the symmetric exchange coupling $\mathcal{J}_s$ to be comparable to the local decay rate and gradually vary the strength of the DM interaction $\mathcal{D}$. Fig.~\ref{figsmf3} (a) demonstrates that as $\mathcal{D}$ increases, the maximal entanglement that can be achieved in the time evolution increases. This is what we expect since more probability flows to the singlet state before significant decoherence occurs, as illustrated in the plots for $G_s(t)$ and $G_t(t)$ in Fig.~\ref{figsmf3} (c) and (d), respectively. The green curves in (c) and (d) indicate that $G_s$ reaches its first peak whereas $G_t$ reaches its first valley after a certain time. For the red curves in (c) and (d), $G_s$ reaches the peak at a later time and the peak is lower than the green curve. Additionally, the entanglement oscillation frequency increases as we increase $\mathcal{D}$, which is understandable as the coherent interaction generally leads to oscillation in the system (i.e., shuffle the information back and forth), as observed in Fig.~\ref{figsmf3} (b), (c), and (d), where all elements oscillate faster. Finally, we observe that the generated entanglement has a shorter lifetime, which is easy to understand since the DM interaction can bring the singlet state (leading to the residual entanglement) that decays slower to the triplet state that decays much faster. This is also evident in Fig.~\ref{figsmf3}(c), where $G_s$ decays much faster as the DM interaction increases.

In the second scenario, we assume that the DM interaction $\mathcal{D}$ is comparable to the local decay rate. Surprisingly, we observe that a increase in the symmetric exchange interaction does not lead to a significant increase in the maximum entanglement, as illustrated in Fig. \ref{figsmf3} (e). This is because the symmetric exchange interaction directly couples $\Re G_{ts}$ to $\Im G_{ts}$, whereas the maximum entanglement depends on the peak that $G_s$ can reach, which is mainly determined by the value of $\mathcal{D}$. However, increasing $\mathcal{J}_s$ does lead to a faster oscillation in all elements, including the entanglement between the two qubits, as we can see in Fig. \ref{figsmf3} (f), (g) and (h).

         \subsection{ Pure classical noise}\label{appendixd5} 
      In Section~\ref{seciv} of the main text, we mention that pure classical noise does not lead to any interesting dynamics. To illustrate this point, we first note that in the absence of any quantum noise, both symmetric exchange and DM interactions are absent, and the decay and absorption rates are equal, since their asymmetry is rooted in the quantum noise. We introduce the notations $\bar{\gamma} \equiv \gamma^\uparrow = \gamma^\downarrow$ for local decay and absorption, and $\bar{\gamma}_{12} \equiv \gamma^\uparrow_{12} = \gamma^\downarrow_{12}$ for the correlated decay and absorption. In this case, the coupled dynamics can be reduced to the following equations:
         \( \al{  &\dot{G}_{11}=-2 \bar{\gamma}G_{11} +(\bar{\gamma}+\bar{\gamma}_{12})G_t +(\bar{\gamma}-\bar{\gamma}_{12})G_s, \\
                     &\dot{G}_{t}=(\bar{\gamma}+\bar{\gamma}_{12}) G_{11} -2(\bar{\gamma}+\bar{\gamma}_{12}) G_t +(\bar{\gamma}+\bar{\gamma}_{12})G_{44}, \\
                     &\dot{G}_s =(\bar{\gamma}-\bar{\gamma}_{12})G_{11}-2(\bar{\gamma}-\bar{\gamma}_{12})G_s +(\bar{\gamma}-\bar{\gamma}_{12})G_{44},\\
                      &\dot{G}_{44}=(\bar{\gamma}+\bar{\gamma}_{12})G_t +(\bar{\gamma}-\bar{\gamma}_{12})G_s-2\bar{\gamma}G_{44},     } \label{D27}  \)
                      and the off-diagonal elements are decoupled from these diagonal elements: $\dot{x}=-2\bar{\gamma}x$ and $\dot{y}=-2\bar{\gamma}y$. Starting from these equations, it is straightforward to verify that the entanglement remains zero if the system is initialized to a product state. To demonstrate the impact of correlated classical noise on the decoherence process, we consider a specific initial state,  the Bell state $\ket{\psi_0}=(\ket{\uparrow\downarrow}+i\ket{\downarrow\uparrow})/\sqrt{2}$. Figure~\ref{figsmf4} displays the entanglement decay for different strengths of correlated classical noise. Although the presence of classical noise modifies the decoherence process slightly, it does not introduce any new features. Therefore, in the main text, we concentrate on the quantum regime where both correlated classical and quantum noise coexist.

\section{Correlated classical and quantum $1/f$ noise}\label{appendixe}
In this section, we provide the detailed derivations for some results presented in Sec.~\ref{secv} in the main text. We first consider the case of purely classical $1/f$ noise and then consider the scenario where the quantum $1/f$ noise is comparable to the classical one.  We assume the correlated noise is comparable to the local noise and is real valued $S_{12}(\omega)=S_{ii}$, as we assumed in the main text. 

\textbf{Correlated classical $1/f$ noise.} In the presence of purely classical $1/f$ noise, the coherent coupling is absent. The decay and absorption rates take the following form from Eq.~\eqref{eqB25}: 
\( \al{    \gamma_{ij}^\downarrow(t)  \!\!&= \!\! \frac{2}{\hbar^2} \!\!\int^\infty_0\!\! \frac{d\omega}{2\pi}  \Big[ S_{ij}^C(\omega) F_s(\omega\!-\!\Omega, t) \!\!+\!\!S^C_{ji}(\omega) F_s(\omega\!+\!\Omega, t)  \Big],  \\
              \gamma_{ij}^\uparrow(t)  \!\!&= \!\! \frac{2}{\hbar^2} \!\!\int^\infty_0\!\! \frac{d\omega}{2\pi}  \Big[ S_{ji}^C(\omega) F_s(\omega\!-\!\Omega, t) \!\!+\!\!S^C_{ij}(\omega) F_s(\omega\!+\!\Omega, t)  \Big] .    } \)
With the assumption that $S_{ij}^C(\omega)=2\pi\sigma^2/|\omega|$, we have the equal (local and correlated) absorption and decay rates [denoted as $ {\gamma}(t)\equiv  \gamma_{ij}^\downarrow(t)=  \gamma_{ij}^\uparrow(t)$], which is given by
\(  \al{  {\gamma}(t)&= \frac{2}{\hbar^2}\int^\infty_{\omega_l} \frac{d\omega}{2\pi} \frac{2\pi \sigma^2}{\omega} \big[ F_s(\omega-\Omega, t) + F_s(\omega+\Omega, t)   \big] \\
                                      & = \frac{4\sigma^2}{\hbar^2\Omega} \big[  \text{Si}(\Omega t ) - \sin(\Omega t) \text{Ci}(\omega_l t)   \big],   }\)
                                      which is the equation~\eqref{eq33} in the main text. One surprising feature is that the above decoherence rate can be temporarily negative.  Here we have introduce the low frequency cutoff $\omega_l$ for the $1/f$ noise. For the purely classical $1/f$ noise, the system is governed by the same set of equations~\eqref{D27} with $\bar{\gamma}=\bar{\gamma}_{12}=\gamma(t)$. First, we can obtain the expression for $x(t)$ and $y(t)$ easily as they are decoupled from other elements: 
                                      \(  x(t)=x(0)e^{-2\Gamma(t)}, \;\;\text{and}\;\;   y(t)=y(0)e^{-2\Gamma(t)},    \)
   where we have introduced $\Gamma(t)=\int^t_0 ds\,\gamma(s)$. We now are interested in how the entanglement decays with the decoherence rate $\gamma(t)$. To be specific, we assume the initial state is a Bell state $\ket{\psi_0}=( \ket{\uparrow\downarrow} +i\ket{\downarrow\uparrow} )/\sqrt{2}$, or equivalently, $G_t(0)=G_s(0)=y(0)=1/2$ (other elements vanish). By using the symmetry between the absorption and decay processes, we conclude that $G_{11}(t)=G_{44}(t)$. In the case of large correlated noise (comparable to local noise), $G_s(t)=1/2$ remains to be a constant. From the fact $\tr \rho=1$, we have the relation $G_t(t)=1/2-2G_{11}$. We can deduce the equation for $G_{11}$: 
   \(  \dot{G}_{11}(t)=-6\gamma(t) G_{11}(t) +\gamma(t),  \)
   which is solved to be
   \(   G_{11}(t)=\int^t_0d\tau \; \Bigg\{\gamma(\tau) \exp[ -6\int^t_\tau ds \, \gamma(s) ] \Bigg\}. \)
 This can be simplified to $G_{11}(t)=[1-\text{exp}[-6\Gamma(t)]]/6$. 
                          One can similarly show that $G_t(t)=1/6+\text{exp}[-6\Gamma(t)]/3$. Then one can evaluate the entanglement according to Eq.~\eqref{D12}, which gives us Eq.~\eqref{eq35} in the main text. When the initial state is a trivial product state, for example $\ket{\uparrow\downarrow}$ (namely, $G_t(0)=G_s(0)=y(0)=1/2$), one can also show that the entanglement remains to be zero with the pure classical $1/f$ noise.

\textbf{Correlated quantum $1/f$ noise.} In the presence of correlated quantum noise $S^Q\approx S^C$, the coherent coupling $\mathcal{J}$ is finite (we assume the spectral density is real), which is evaluated to be: 
\( \al{\mathcal{J}(t)& =\frac{2}{\hbar^2}\int^\infty_0 \frac{d\omega}{2\pi} \frac{2\pi \sigma^2}{\omega} \big[  F_c(\omega-\Omega, t) + F_c(\omega+\Omega, t)  \big] \\
                             &=- \frac{2\pi \sigma^2}{\hbar^2 \Omega} \sin\Omega t, }\)
where we have taken the principle value of the integral. This is Eq.~\eqref{eq39} in the main text. As we discussed in the main text, we approximate the absorption rate with zero and the decay rate is evaluated to be Eq.~\eqref{eq37}. In this case, the two-qubit dynamics is governed by the same set of equations in Eq.~\eqref{eqd14} but with $\mathcal{J}_s=\mathcal{J}(t), \gamma^\downarrow=\gamma_{12}^\downarrow$ and $\mathcal{D}=0.$ We again consider two initial states: one is the Bell state $\ket{\psi_0}$ and the other one is the product state $\ket{\uparrow\downarrow}.$ In both cases,  we can show that $G_s(t)=1/2$ and $G_t(t)=e^{-2\Gamma^\downarrow(t)}/2$ with $\Gamma^\downarrow(t)=\int^t_0ds\gamma^\downarrow(s)$. For the dynamics of $\Re G_{ts}$ and $\Im G_{ts}$, we introduce $X(t)\equiv xe^{\Gamma^\downarrow(t)}$ and  $Y(t)\equiv y e^{\Gamma^\downarrow(t)}$. One can show that they are described by the following equations: 
\(  \dv{}{t}\mqty[X(t) \\ Y(t)] =\mqty[0 & 2\mathcal{J}(t)  \\ -2\mathcal{J}(t)  & 0 ] \mqty[X(t) \\ Y(t)] ,   \)
from which we can obtain: $[X(t), Y(t)]^T=U(t) [X(0), Y(0)]^T$ with the rotation matrix: 
\( U(t) =\mqty[ \cos \Phi(t)  & -\sin\Phi(t) \\ \sin\Phi(t)  & \cos \Phi(t)  ],    \)
where $\Phi(t)=\int^t_0ds\, \mathcal{J}(t)$. When the initial state is $\ket{\uparrow\downarrow}$, we have initial condition $X(0)=1/2$ and $Y(0)=0$ and obtain: 
\( y(t) =\frac{ \sin\Phi(t)   }{2} \exp[-\Gamma^\downarrow(t)]. \)
In this case, the entanglement is given by Eq.~\eqref{eq42} in the main text. When the initial state is the Bell state $\ket{\psi_0}$, we have $X(0)=0$ and $Y(0)=1/2$, which gives us 
\( y(t) =\frac{ \cos\Phi(t)   }{2} \exp[-\Gamma^\downarrow(t)]. \)
This leads to the entanglement in Eq.~\eqref{eq40}. So far, we have focused on the quantum regime, and it was shown in the main text that the final entanglement in this case is 1/2. However, we also wish to examine the final entanglement as a function of temperature, or equivalently, the ratio between the quantum noise and classical noise, by invoking the following relation
\( \cosh\beta\hbar \Omega = \frac{[S^C(\Omega)  ]^2 +[S^Q(\Omega) ]^2  }{[S^C(\Omega)  ]^2 -[S^Q(\Omega) ]^2 } . \)
To this end, we set $\dot{\rho}=0$ and $t\rightarrow \infty$ [then the dynamics is governed by Eq.~\eqref{D3} with all coefficients being constant in this limit]. We still assume the correlated noise is comparable to the local noise (otherwise, one can show the entanglement will eventually decay to zero). Then we can show that $G_s=1/2$ and $G_{11}=\alpha G_t=\alpha^2G_{44}$ with $2G_{44}=(1+\alpha+\alpha^2)^{-1}$ and $\alpha=e^{-\beta\hbar \Omega}$ and other off-diagonal elements are zero. From these results, we can obtain Eq.~\eqref{eq41} in the main text.


\begin{thebibliography}{96}
\bibitem [{cor()}]{correlated_noise_phase}%
  \BibitemOpen
  \href@noop {} {}\bibinfo {note} {The phase $\theta$ has a frequency
  dependence in general whose exact form depends on the details of the
  environment and the coupling. In this work, we assume the phase to be a
  constant for simplicity.}\BibitemShut {Stop}%
\end{thebibliography}
%

\end{document}